\documentclass[prb,floatfix,twocolumn,showpacs,amsmath,amssymb,superscriptaddress,titlepage]{revtex4-1}
\usepackage{graphicx}
\usepackage{epstopdf}
\usepackage{epsfig}
\usepackage{dcolumn}
\usepackage{bm}
\usepackage{color}
\usepackage{subfigure}
\usepackage{wrapfig}
\usepackage{rotate,color,url}
\usepackage{longtable}
\usepackage{time}
\usepackage[pagebackref=false,colorlinks,linkcolor=blue,citecolor=blue,urlcolor=magenta]{hyperref}

\begin{document}
\newcommand {\be}{\begin{equation}}
\newcommand {\ee}{\end{equation}}
\newcommand {\bea}{\begin{array}}
\newcommand {\cl}{\centerline}
\newcommand {\eea}{\end{array}}
\newcommand {\pa}{\partial}
\newcommand {\al}{\alpha}
\newcommand {\de}{\delta}
\newcommand {\ta}{\tau}
\newcommand {\ga}{\gamma}
\newcommand {\ep}{\epsilon}
\newcommand {\si}{\sigma}
\newcommand{\up}{\uparrow}
\newcommand{\down}{\downarrow}

\title{  Quantum phase diagram of  $J_1-J_2$  Heisenberg $S=1/2$ antiferromagnet in honeycomb lattice: a modified spin wave study}

\author{Elaheh Ghorbani}
\email{el.ghorbani@gmail.com}
\affiliation{Department of Physics, Isfahan University of Technology, Isfahan 84156-83111, Iran}

\author{Farhad Shahbazi}
\email{shahbazi@cc.iut.ac.ir}
\affiliation{Department of Physics, Isfahan University of Technology, Isfahan 84156-83111, Iran}
\affiliation{School of Physics, Institute for Research in Fundamental Sciences (IPM), Tehran 19395-5531, Iran}

\author{Hamid Mosadeq}
\email{h-mosadeq@ph.iut.ac.ir}
\affiliation{Department of Physics, Shahrekord University, Shahrekord, Iran}
\affiliation{School of Physics, Institute for Research in Fundamental Sciences (IPM), Tehran 19395-5531, Iran}

\date{\today}

\begin{abstract}
Using modified spin wave (MSW) method, we study the $J_1-J_2$ Heisenberg model with first and second neighbor antiferromagnetic exchange interactions. For symmetric $S=1/2$ model,  with the same couplings for all the equivalent neighbors, we find three phase in terms of frustration parameter ${\bar\alpha=J_2/J_1}$: (1) a commensurate collinear ordering  with staggered magnetization (N{\'e}el.I state) for $0\leq{\bar\alpha}\lesssim 0.207$ , (2) a magnetically gapped disordered state for $0.207\lesssim{\bar\alpha}\lesssim 0.369$,  preserving all the symmetries of the Hamiltonian and lattice, hence by definition is  a quantum spin liquid (QSL) state and (3) a commensurate collinear ordering in which two out of three nearest neighbor magnetizations are antiparallel and the remaining pair are parallel (N{\'e}el.II state), for  $0.396\lesssim{\bar\alpha}\leq 1$. 
We also explore the phase diagram of distorted $J_1-J_2$ model with $S=1/2$. Distortion is introduced as an inequality of one nearest neighbor coupling with the other two. This yields a richer   phase diagram  by the appearance of a new gapped QSL, a gapless QSL and also a valence bond crystal (VBC) phase in addition to the previously three phases found for undistorted model. 
   
\end{abstract}

\pacs{
75.10.Jm	
75.10.Kt	
75.50.Ee   
} 
\maketitle
\section{Introduction \label{int}}
Recent synthesis of compounds consisting of transition metal-oxide layers with honeycomb structure, has drawn the attentions to the magnetic properties of the spin models in honeycomb lattice. Three experimental realizations of honeycomb magnetic materials are InCu$_{2/3}$V$_{1/3}$O$_3$~\cite{InCuVO}, Cu$_3$Ni$_2$SbO$_6$~\cite{CuNiSbO} and Bi$_{3}$Mn$_{4}$O$_{12}$(NO$_{3}$) (BMNO)~\cite{bmno1}.   Cu$^{+2}$ ions with $S=1/2$  in the first, Ni$^+$ ions with $S=1$ in the second and Mn$^{+4}$ ions with $S=3/2$ in the third compound reside on the lattice points of weakly coupled honeycomb layers. InCu$_{2/3}$V$_{1/3}$O$_3$ develops  antiferromagnetic (AF) ordering below $\sim 20$K~\cite{InCuVO2}. However, for BMNO the magnetic susceptibility as well as specific heat measurements show no sign of  magnetic ordering  down to $T=0.4$K, in spite of the high Curie-Weiss temperature $T_{\rm CW}\approx -257 $K~\cite{bmno1}. 

On the theoretical front, the large scale Quantum Monte Carlo (QMC) simulation of the half-filled Hubbard
model on the honeycomb lattice, proposes a gapped  quantum spin liquid (QSL) phase (a magnetically disordered state preserving all the symmetries of the Hamiltonian and the lattice) for intermediate values of on-site coulomb interaction between the   AF-Mott insulating  and the  semi-metallic phases~\cite{meng}.   Although,  later QMC simulations on larger lattice sizes  refuted the existence of such a QSL phase~\cite{debate1,debate2,debate3}, nevertheless, many  researches   were devoted  to the study of AF spin models in  honeycomb structure~\cite{katsura,QMC,nlsigma,sw,series,fouet,takano,noorbakhsh,kawamura,Aron2010,sd2,mosadeq,pvb2,pvb3,pvb4,pvb5,pvb6,pvb7,pvb8,SB-lamas1,SB-wang,SF-lu,VMC-sondhi,SB-lamas2,SB-china,eps,ring1,ring2,zare,bishop,Ciolo,DMRG-s1,bishop-s1}.

Since honeycomb lattice is  bipartite, Heisenberg model with nearest neighbor AF interactions in the this lattice is not frustrated and develops long-range N{\'e}el ordering. However, enhanced quantum fluctuations, due to the small coordination number ($Z=3$),  reduce the staggered magnetization  by about half of its classical value~\cite{QMC,series,sw,noorbakhsh}. Therefore, the expectation for realization of a QSL phase in honeycomb based magnets, requires the introduction of frustrating exchange interactions. The simplest model incorporating frustration effects on the honeycomb lattice is $J_{1}-J_{2}$ Heisenberg model, where $J_{1}>0$ and $J_{2}>0$ are nearest and next to nearest neighbor AF exchange interactions, respectively. The classical phase diagram of this model,  studied by Katsura {\em et al}~\cite{katsura},  shows  that the N{\'e}el ordered phase is stable for $J_2/J_1< 1/6$. However, for $1/6<J_2/J_1< 1/2$ the classical ground state becomes infinitely degenerate and can be  characterized by a manifold of spiral wave vectors. Okumura {\em et al}, used the low temperature expansion and Monte Carlo (MC) simulation, to show that the such a large ground state degeneracy can be lifted by thermal fluctuations in such  a way that a broken symmetry state, with three-fold ($C_{3}$) symmetry of the honeycomb lattice, would be selected~\cite{kawamura}. In the vicinity of AF phase boundary ($J_2/J_1\sim 1/6$), the energy scale associated  with such a  thermal order by disorder mechanism becomes  extremely small, leading to exotic spin liquid behaviors, whereby the spin structure factor would  have different pattern in  comparing with the paramagnetic phase~\cite{kawamura}. 

Order by disorder mechanism driven by quantum fluctuations  has been studied by Mulder {\em et al}. They showed that the spin wave corrections lower the energy of some states with  particular  incommensurate wave vectors in the ground state manifold, for the classically degenerate region $1/6<J_2/J_1<1/2$~\cite{Aron2010}. They also argued that for $S=1/2$, over a wide range of $J_2/J_1$ in the frustrated region, strong quantum fluctuations can melt  this spiral ordering  into a  valence bond solid (VBS) with staggered dimerized  ordering, which breaks the $C_{3}$ rotational symmetry of the lattice while preserving  its translational symmetry~\cite{Aron2010}. Such a  nematic ordering has already been proposed in exact diagonalization (ED) calculations~\cite{fouet} and also by non-linear sigma model formulation~\cite{takano}.  
ED calculations in both $S_z=0$, and nearest neighbour  valence bond (NNVB) basis, show that NNVB basis provides a very good description of the ground state for $0.2\lesssim J_2/J_1 \lesssim 0.3$~\cite{mosadeq, pvb2}. Furthermore,  analysis of the ground state properties by defining appropriate structure factors, suggests a plaquette valence bond solid (PVBS) ground state for $0.2\lesssim J_2/J_1\lesssim 0.35$ which transforms to a  valence bond solid (VBS) state with staggered dimerization  at $J_2/J_1\approx 0.35$~\cite{mosadeq, pvb2}. The existence of plaquette valence bond solid has  been verified  by different methods, such as functional renormalization group ~\cite{pvb3}, coupled cluster method (CCM)~\cite{pvb4,pvb5}, mean-field plaquette valence bond theory~\cite{pvb6,zare}
 and density matrix renormalization group (DMRG)~\cite{pvb7,pvb8}. However, other methods such as Schwinger boson mean-field approach~\cite{SB-lamas1,SB-wang,SB-lamas2,SB-china}, Schwinger fermion mean-field theory~\cite{SF-lu} and variational Monte Carlo~\cite{VMC-sondhi} propose a $Z_2$ quantum spin liquid (QSL) for the disordered region.

In this work we use the the modified spin wave (MSW) theory to  study both  symmetric  and  distorted $J_1-J_2$ Heisenberg AF with $S=1/2$ model in the honeycomb lattice. 
This paper is organized a follows: the  model Hamiltonian and the modified spin wave method are introduced in section \ref{msw}. The MSW phase diagram of symmetric and distorted model is discussed in sections \ref{j1j2-iso}  and \ref{j1j2-dis}. Section \ref{conclusion}  is devoted to conclusion. 
\section{Model Hamiltonian and Modified Spin-Wave (MSW) formalism}
\label{msw}
\begin{figure}[b] 
\centering 
\includegraphics[width=0.7\columnwidth]{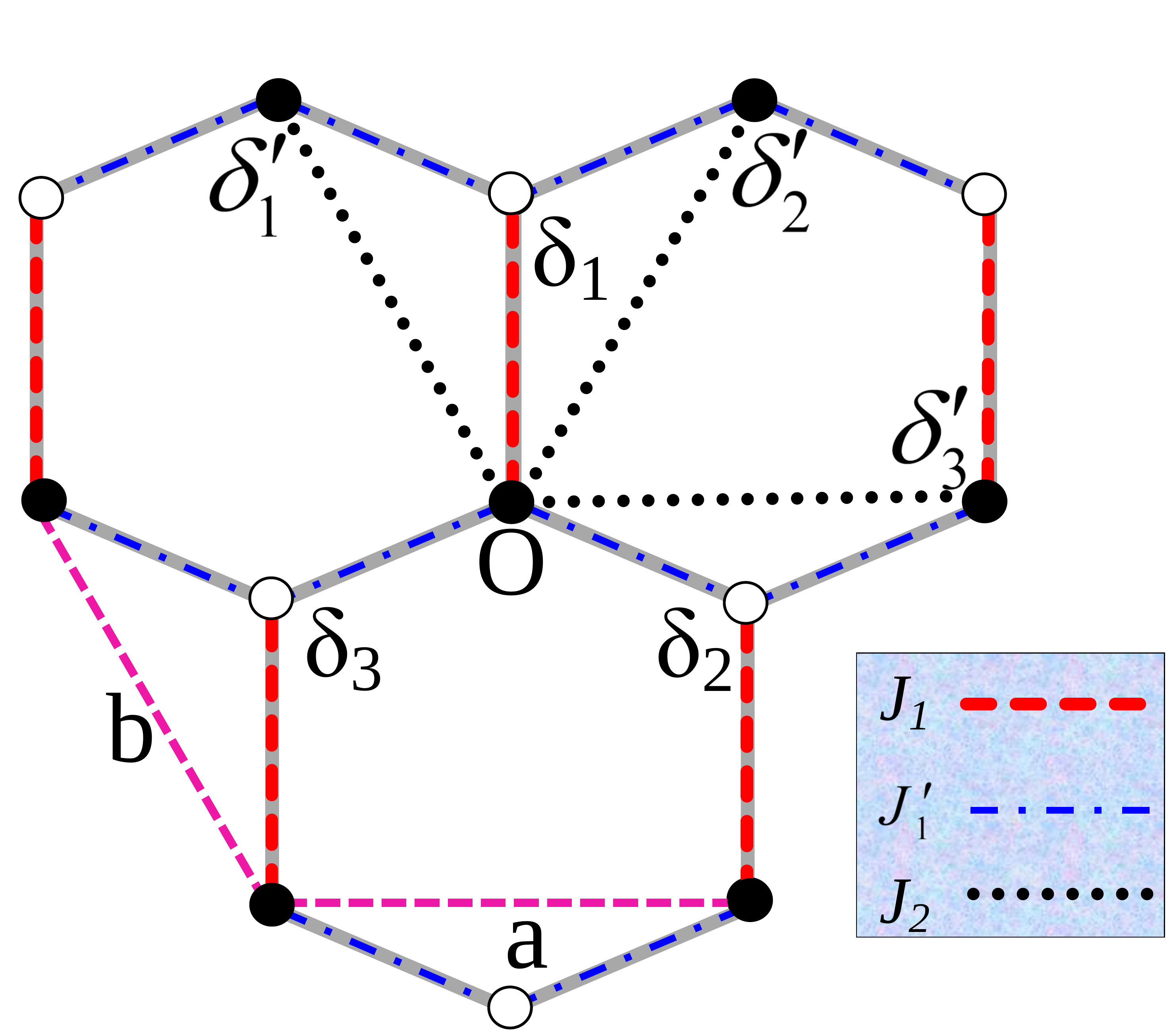} 
\caption{(Color online) Honeycomb lattice with basis vectors $\bf \delta_{0}=0$ and $\bf \delta_{1}=\frac{a}{\sqrt{3}}{\hat j}$, and primitive vectors ${{\bf a}=a{\hat i}}$ and ${{\bf b}=-\frac{a}{2}{\hat i}+\frac{\sqrt{3}a}{2}{\hat j}}$. Black and white circles indicate A and B sublattices, respectively. $\delta_1$, $\delta_2$ and $\delta_3$ denote the nearest neighbors and $\delta_1^\prime$, $\delta_2^\prime$ and $\delta_3^\prime$ represent  the next nearest neighbors.} 
\label{honeycomb} 
\end{figure}

The  $J_1-J_2$ Heisenberg AF Hamiltonian is defined by, 
\be 
H={ \frac{1}{2}}\sum_{nn}J_{ij}{\bf S}_i\cdot{\bf S}_j 
+{ \frac{J_2}{2}}\sum_{nnn}{\bf S}_i\cdot{\bf S}_j, 
\label{j1j2-distort} 
\ee 
in which $nn$ and $nnn$ denote nearest and next nearest neighbors, respectively, and the exchange coupling $J_{ij}>0$ and $J_2>0$, denote the first and second neighbor couplings. Here we consider the case where the nearest neighbor couplings are equal to $J_1$ for the bond denoted by the vector $\delta_1$ and $J'_1$ for  the bonds denoted by $\delta_2$ and $\delta_3$ (see Fig. \ref{honeycomb}).  Now we redefine the couplings as follows

\be
J'_1=\bar{J_1}(1-\bar{\delta}),\ J_1=\bar{J_1}(1+2\bar{\delta}),\ \bar{\alpha}=J_2/\bar{J_1},
\label{j1j2-def}
\ee
where the dimensionless quantities ${\bar\delta}$ and ${\bar\alpha}$ denote the distortion and frustration, respectively.

Now, we  give a brief  introduction  to the formalism of MSW theory in a bipartite lattice, and then apply it to the Hamiltonian \eqref{j1j2-distort}. 
MSW was introduced by Takahashi~\cite{Takahashi} and its basic assumption  is that the ground state of spin Hamiltonian in the classical limit ($S\rightarrow\infty$), is long-range ordered. It has been shown that minimum energy condition for the classical $ J_{1}-J_{2} $ Heisenberg model, gives rise to planar states~\cite{katsura,fouet}. 
Hence, the translational invariance requires that the ordered ground state is characterized by a planar wave-vector ${\bf Q}$. Under this assumption, it is convenient to rotate the coordinate axes $(x, y, z)$ locally to ($\eta_i$, $\zeta_i$, $\xi_i$) at each site $i$, in such a way that  $ \zeta_i $ represents the local symmetry breaking axis. For this purpose, we introduce the following spin transformations for the honeycomb lattice which contains two lattice points per unit cell 
\begin{eqnarray}
 S_{i,\gamma}^x=&-&\sin\left({\bf Q}\cdot{\bf  r}_{i}+(\gamma-1)\phi)\right) S_{i,\gamma}^\eta\nonumber\\ 
&+&\cos\left({\bf Q}\cdot{\bf r}_{i}+(\gamma-1)\phi)\right) S_{i,\gamma}^\zeta \nonumber\\ 
 S_{i,\gamma}^y=& &\cos\left({\bf Q}\cdot{\bf  r}_{i}+(\gamma-1)\phi)\right) S_{i,\gamma}^\eta\nonumber \\
&+&\sin\left({\bf Q}\cdot{\bf  r}_{i}+(\gamma-1)\phi)\right) S_{i,\gamma}^\zeta \nonumber\\
 S_{i,\gamma}^z=&-&S_{i,\gamma}^\xi,
\label{rotate}
\end{eqnarray}
 where ${\bf r}_{i}$ denotes the position of each spin, $\gamma=1,2$ refers to the two lattice points (A, B sublattices) in the unit cell identified by the  vectors $\delta_{0}=0$ and $\delta_{1}=\frac{a}{\sqrt{3}}{\hat j}$ ( see Fig. \ref{honeycomb}) and the angle $\phi$ denotes the relative rotation of the symmetry breaking axes within a unit cell. Unlike ordinary spin-wave theory, we do not make any assumption on the ordering vector ${\bf Q}$ which may differ from the classical ordering wave vector. 

Applying the transformations \eqref{rotate} to the Hamiltonian \eqref{j1j2-distort}, we find  
\begin{widetext} 
\begin{eqnarray} 
H&={\frac{1}{2}}&\sum_{nn}J_{ij}\left[({S}_{i,1}^\eta S_{j,2}^\eta+S_{i,1}^\zeta S_{j,2}^\zeta)\cos[{\bf Q}\cdot({\bf r}_i-{\bf r}_j)+\phi]
+(-S_{i,1}^\eta S_{j,2}^\zeta+S_{i,1}^\zeta S_{j,2}^\eta)\sin[{\bf Q}\cdot({\bf r}_i-{\bf r}_j+\phi]+S_{i,1}^\xi S_{j,2}^\xi\right]\nonumber\\ 
&+&{\frac{J_2}{2}}\sum_{nnn}\sum_{\gamma=1}^{2}\left[(S_{i,\gamma}^\eta S_{j,\gamma}^\eta+S_{i,\gamma}^\zeta S_{j,\gamma}^\zeta)\cos[{\bf Q}\cdot({\bf r}_i-{\bf r}_j) ]
+(-S_{i,\gamma}^\eta S_{j,\gamma}^\zeta+S_{i,\gamma}^\zeta S_{j,\gamma}^\eta)\sin[{\bf Q}(\cdot{\bf r}_i-{\bf r}_j)]+S_{i,\gamma}^\xi S_{j,\gamma}^\xi\right].
\label{rotateH} 
\end{eqnarray} 
\end{widetext} 
We use Dyson-Maleev (DM) transformations to obtain a bosonic representation of the spin Hamiltonian. 
For a bipartite lattice, like the honeycomb lattice, DM transformations are given by
\begin{equation} 
\begin{split} 
S_{i,1}^-=\frac{1}{\sqrt{2S}}(2S-a_i^\dag a_i)a_i,\ \ \ &S_{i,2}^-=\frac{1}{\sqrt{2S}}(2S-b_i^\dag b_i)b_i, \\ 
S_{i,1}^+=\sqrt{2S}a_i^\dag, \ \ \ \ \ \ \ \ \ \ \ \ \ \ \ \ \ \  &S_{i,2}^+=\sqrt{2S}b_i^\dag, \\ 
S_{i,1}^\zeta=S-a_i^\dag a_i,\ \ \ \ \ \ \ \ \ \ \ \ \ \ \ \ &S_{i,2}^\zeta=S-b_i^\dag b_i, 
\end{split} 
\label{DM} 
\end{equation} 
in which $a,b$ represent the bosonic operators in A and B sublattices, respectively, and $S$ is the value of the spins. In the above transformations the quantization axes are taken to be the local $\zeta$ axes and $S_i^\pm\equiv S_i^\eta\pm i S_i^\xi$. The commutation relations $[S_i^\alpha,S_j^\beta]=i\epsilon_{\alpha\beta\gamma}S_i^\gamma \delta_{ij}$ are satisfied by the bosonic algebra between $a$ and $b$ operators, i.e, $[a_i,a_{i^\prime}^\dag]=\delta_{ii^\prime}$, $[b_j,b_{j^\prime}^\dag]=\delta_{jj^\prime}, [a_i,a_{i^\prime}]=[b_i,b_{i^\prime}]=[a_i,b_{i^\prime}]=0$. 
Substituting the transformations \eqref{DM} into the Hamiltonian \eqref{rotate}, one finds the following bosonic Hamiltonian 
\begin{widetext} 
\begin{equation} 
\begin{split} 
H=&\frac{1}{4}\sum_{nn}J_{ij}\{[2S(a_i^\dag b_j+a_i b_j^\dag)-a_i^\dag b_j^\dag b_jb_j-a_i^\dag a_ia_ib_j^\dag](1+\cos({\bf Q}\cdot{\bf r}_{ij}+\phi))\\ 
&+[-2S(a_i^\dag b_j^\dag+a_i b_j)+a_i b_j^\dag b_jb_j+a_i^\dag a_i a_ib_j](1-\cos({\bf Q}\cdot{\bf r}_{ij}+\phi))\\ 
&+4[S^2-S(a_i^\dag a_i+b_j^\dag b_j)+a_i^\dag a_i b_j^\dag b_j]\cos({\bf Q}\cdot{\bf r}_{ij}+\phi)\}\\ 
&+\frac{J_2}{4}\sum_{nnn}[2S(a_i^\dag a_j+a_i a_j^\dag)-a_i^\dag a_j^\dag a_ja_j-a_i^\dag a_ia_ia_j^\dag](1+\cos({\bf Q}\cdot{\bf r}_{ij}))\\ 
&+[-2S(a_i^\dag a_j^\dag+a_i a_j)+a_i a_j^\dag a_ja_j+a_i^\dag a_i a_ia_j](1-\cos({\bf Q}\cdot{\bf r}_{ij}))\\ 
&+4[S^2-S(a_i^\dag a_i+a_j^\dag a_j)+a_i^\dag a_i a_j^\dag a_j]\cos({\bf Q}\cdot{\bf r}_{ij})\\ 
&+\frac{J_2}{4}\sum_{nnn}[2S(b_i^\dag b_j+b_i b_j^\dag)-b_i^\dag b_j^\dag b_jb_j-b_i^\dag b_ib_ib_j^\dag](1+\cos({\bf Q}\cdot{\bf r}_{ij}))\\ 
&+[-2S(b_i^\dag b_j^\dag+b_i b_j)+b_i b_j^\dag b_jb_j+b_i^\dag b_i b_ib_j](1-\cos({\bf Q}\cdot{\bf r}_{ij}))\\ 
&+4[S^2-S(b_i^\dag b_i+b_j^\dag b_j)+b_i^\dag b_i b_j^\dag b_j]\cos({\bf Q}\cdot{\bf r}_{ij}), 
\end{split} 
\label{DMH} 
\end{equation} 
\end{widetext} 
where ${\bf r}_{ij}={\bf r}_i-{\bf r}_j$ is equal to $\delta_1, \delta_2, \delta_3$ for the nearest neighbors and $\pm\delta'_1, \pm\delta'_2, \pm\delta'_3$ for the next to nearest neighbors (Fig. \ref{honeycomb}). Now, we  use mean field theory to find an expression for the expectation value of the Hamiltonian \eqref{DMH}, i.e. $E=\langle H\rangle$. For this purpose, we use the Wick's theorem  to calculate the expectation value of the quartic terms, 
hence we find 
\begin{widetext} 
\begin{equation} 
\begin{split} 
E=& -\frac{N}{2}\sum_{\delta}J(\delta)\{[S+\frac{1}{2}-f(0)+g({\delta})]^2(1-\cos({\bf{Q}}\cdot{{\delta}}+\phi))
-[S+\frac{1}{2}-f(0)+f({\delta})]^2(1+\cos(\bf{Q}\cdot\delta+\phi))\}\\ 
&-\frac{J_2}{2}\sum_{\delta'}\{[S+\frac{1}{2}-f(0)+g(\delta')]^2(1-\cos({\bf{Q}}\cdot{\delta'}))
-[S+\frac{N}{2}-f(0)+f(\delta')]^2(1+\cos({\bf{Q}}\cdot{\delta'}))\}, 
\end{split} 
\label{finalH} 
\end{equation} 
\end{widetext} 
in which $\delta$ and $\delta'$ denote the first and second neighbors, respectively,  $J(\delta_1)=J_1$, $J(\delta_2)=J(\delta_3)=J'_1$ and  $N$ is the number of sites. Functions $f$ and $g$ denote the expectation value of hopping and  pairing  of  DM bosons  defined as 
\begin{equation} 
\begin{split} 
\langle a_i^\dag b_j\rangle=&\langle a_i b_j^\dag\rangle\equiv f(\delta),\
\langle a_i^\dag a_j\rangle = \langle b_i^\dag b_j \rangle\equiv f(\delta')-\frac{1}{2}\delta_{ij},\\ 
\langle a_i^\dag a_j^\dag\rangle=&\langle a_i a_j\rangle\equiv g(\delta'),\
\langle b_i^\dag b_j^\dag \rangle=\langle b_i b_{j}\rangle\equiv g(\delta'),\\ 
\langle a_i^\dag b_{j}^\dag\rangle=&\langle a_i b_{j}\rangle\equiv g(\delta),
\end{split} 
\label{relation} 
\end{equation} 
 with $f(0)=f({\delta=0})$.

 Then using equations \eqref{finalH} and \eqref{relation}, one finds the following expression for the ground state energy per site, $E_0=E/N$, in mean field approximation

\begin{eqnarray} 
&&E_0=\epsilon_0+\epsilon_1\cos(\phi)\nonumber\\ 
&&+\epsilon'_{1}[\cos(\frac{Q_x}{2}+\frac{\sqrt{3}Q_y}{2}+\phi)+\cos(-\frac{Q_x}{2}+\frac{\sqrt{3}Q_y}{2}+\phi)]\nonumber\\ 
&&+\epsilon_2[\cos(Q_x)+\cos(\frac{Q_x}{2}+\frac{\sqrt{3}Q_y}{2})
+\cos(-\frac{Q_x}{2}+\frac{\sqrt{3}Q_y}{2})],\nonumber\\ 
\label{Ehoney} 
\end{eqnarray} 
where 

\begin{eqnarray} 
\epsilon_0&=&
{ \sum _{\delta} \frac{J(\delta)}{2} (f(\delta)^2-g(\delta)^2) }\nonumber\\
&+&
{\frac{ J_2}{2}\sum_{\delta'} (f(\delta')^2-g(\delta')^2)},\nonumber\\ 
\epsilon_1&=&\frac{J_{1}}{2}(f(\delta_1)^2+g(\delta_1)^2), \nonumber \\
\epsilon'_1&=&{\frac{J'_{1}}{4} \sum _{i=2}^{3}(f(\delta_i)^2+g(\delta_i)^2)},\nonumber\\ 
\epsilon_2&=&{ \frac{J_{2}}{12} \sum _{\delta'}(f(\delta')^2+g(\delta')^2)}.
\label{e0e1e2} 
\end{eqnarray} 

First step in MSW procedure is to minimize the  energy \eqref{Ehoney}  with respect to the ordering vector $\bf{Q}$. This incorporates  the competition between states with LRO at different ordering vectors $\bf{Q}$ which may  not necessary be stable  at the classical level~\cite{Ting}.
Next step  is to minimize $E_0$ with respect to $f_{ij}$ and $g_{ij}$. In the absence of external field, this  minimization is done under the constraint that the expectation value of spins along the local quantization axes vanishes. The constraint, $\langle S_i^\zeta\rangle=0$, introduced by Takahashi~\cite{Takahashi}, to keep the number of DM bosons per site less than $2S$ ($n<2S$).
\begin{eqnarray} 
 \langle S_i^\zeta\rangle&=&-S+\langle a_i^\dag a_i\rangle=-S+\langle b_i^\dag b_i\rangle\nonumber\\ 
 &=&-S-\frac{1}{2}+f(0)= 0.
\label{constraint} 
\end{eqnarray} 
The Takahashi's constraint reduces the Hilbert space dimension available to the DM bosons by reducing their average density to $S$. In a bipartite lattice, one can in fact show a significant reduction of the Hilbert space dimension from $2^N$ to $\frac{4}{\pi}\frac{2^N}{N}$ for $S=1/2$~\cite{dotsen}.

For a given ordering wave vector ${\bf Q}$ and rotation $\phi$, an appropriate set of Bogoliubov transformations are defined, in terms of which the Hamiltonian equation \eqref{DMH} can be diagonalized in mean field approximation. Moreover, the quantities $f$ and $g$ defined by equation \eqref{relation} can be parameterized in terms of the coefficients of the Bogoliubov transformations, allowing us to minimize the total energy with respect to these coefficients, under the Takahashi's constraint \eqref{constraint}.  To satisfy the Takahashi's constraint we need to introduce a Lagrange  multiplier $\mu$ which plays the role of chemical potential for the DM bosons. In bosonic language, a magnetically ordered state can be translated to a   Bose-Einstein condensate  (BEC), for which $\mu=0$~\cite{cirac}. For the magnetic disordered states  the spontaneous  magnetization is zero, hence there is no reason for vanishing of the chemical potential. In this case $\mu$ has to be calculated self-consistently to give the gap of the magnon dispersion.  MSW gives  a set of self-consistent equations for  $g$ and $f$,  whose outputs are the ground state energy, magnon energy spectrum, magnetization and spin-spin correlations.  The details of this procedure are given in Appendices  \ref{app.1} and \ref{app.2}.
In the next section we  apply MSW theory to the symmetric $J_1-J_2$ model. 

\section{MSW phase diagram of symmetric $J_1-J_2$ model}
\label{j1j2-iso}

\begin{figure}[t] 
\centering 
\includegraphics[width=\columnwidth]{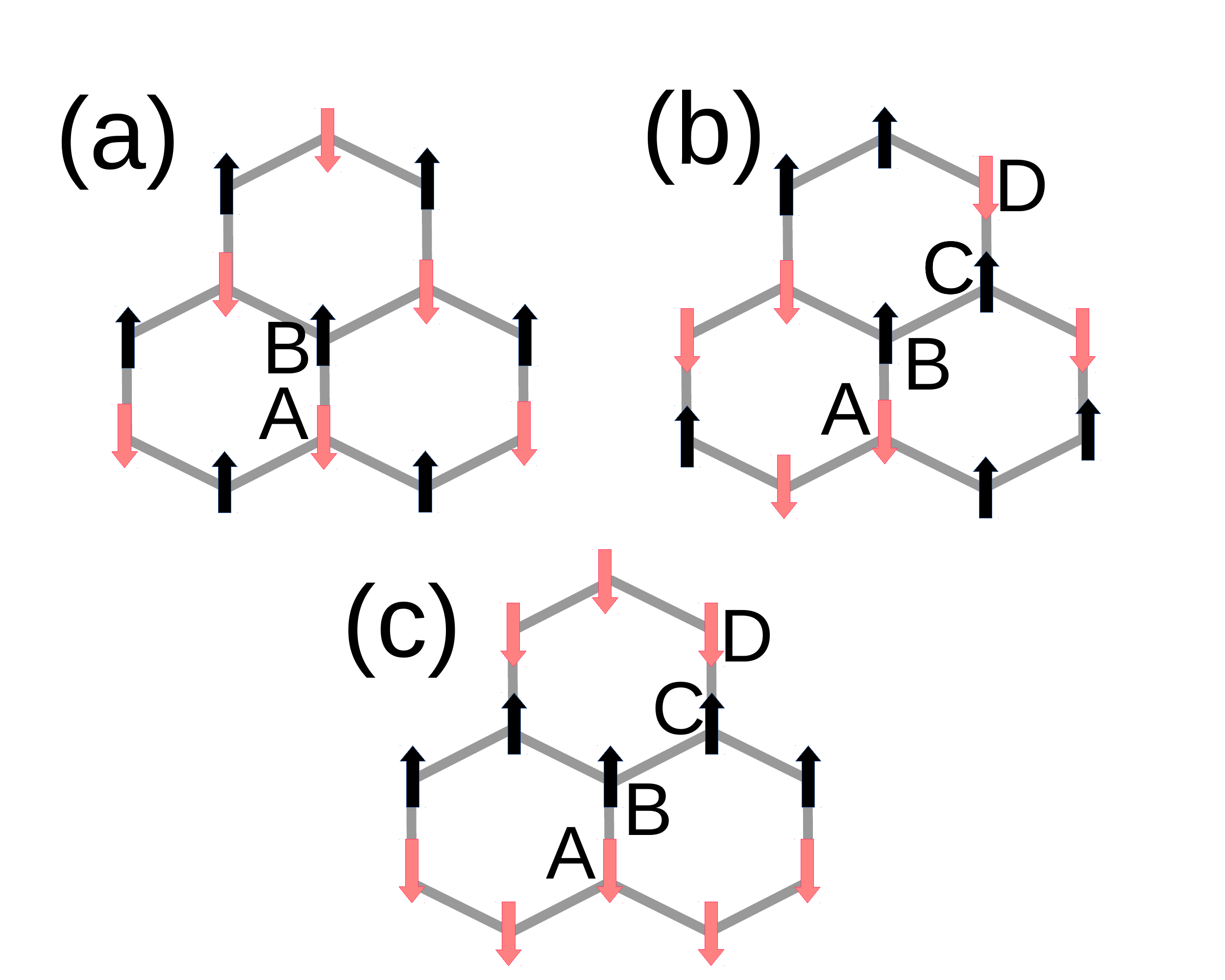} 
\caption{(Color online) Schematic representation of (a) N{\'e}el.I spin configuration comprised of two magnetic sublattices  A and B,  (b) N{\'e}el.II and (c) N{\'e}el.III states consisting of  four magnetic sublattices A, B, C and D. } 
\label{ordered} 
\end{figure}   

 For the symmetric model $J_1=J'_1$, minimizing  the total energy \eqref{Ehoney} with respect to $Q_x$, $Q_y$ and $\phi$, gives rise to numerous commensurate and incommensurate solutions. The commensurate minima are achieved by a two-sublattice  collinear ordering, given by ${\bf Q}=(0,0)$; $\phi=\pi$ (N{\'e}el.I), and  two types of four-sublattices collinear ordering with ${\bf Q}=(\pi,\frac{\pi}{\sqrt{3}})$; $\phi=\pi$ (N{\'e}el.II) and ${\bf Q}=(0,\frac{2\pi}{\sqrt{3}})$; $\phi=0$ (N{\'e}el.III). The schematic spin configurations  in these  states are illustrated  in figure \ref{ordered}. The incommensurate solutions are given by  the spiral states ${\bf Q}=(2\cos^{-1}(\pm \frac{1}{2}\frac{3\epsilon_1\pm \epsilon_2}{\epsilon_2}),0\ {\rm or}\ \frac{2\pi}{3})$; $\phi=0\ {\rm or}\ \pi$ , and ${\bf Q}=(0,\frac{2}{\sqrt{3}}(\sin^{-1}(\frac{1}{2}\sin(\phi))-\phi)$ ; $\phi=\cos^{-1}(\frac{\epsilon_2}{\epsilon_1}-\frac{3\epsilon_1}{4\epsilon_2})+\pi$, where $\epsilon_1$ and $\epsilon_2$ are given by equation \eqref{e0e1e2}. 
 

\begin{figure}[t] 
\begin{center}
\includegraphics[trim = 7mm 0mm 0mm 0mm, clip,width=\columnwidth]{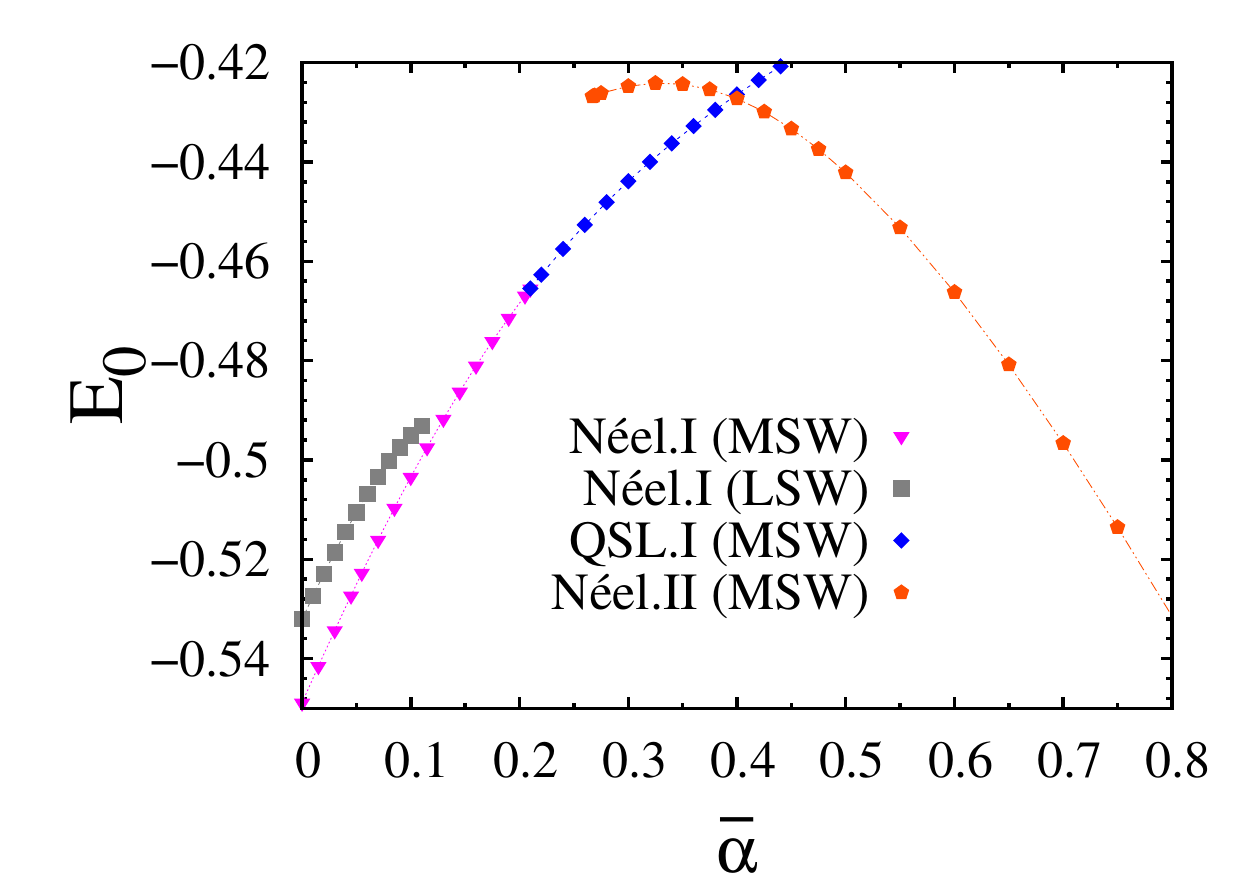} 
\end{center}
\caption{(Color online) Ground state energy per site (in unites of $J_1$) of  
symmetric $J_1-J_2$ model versus the frustration parameter ${\bar\alpha}$ for $S=1/2$.  The ground state is N{\'e}el.I ordered for  $0.0\leq {\bar\alpha}\lesssim0.207$ (pink triangles) , quantum spin liquid with N{\'e}el.I type symmetry (QSL.I) for $0.207\lesssim {\bar\alpha}\lesssim0.396$ (blue rhomboids) and N{\'e}el.II ordered for $0.396\lesssim {\bar\alpha}\leq1.0$ (red diamonds).   The grey squares show the ground state energies per site, obtained in linear spin wave (LSW) approximation in the N{\'e}el.I phase.  } 
\label{energy} 
\end{figure} 
\begin{figure}[t] 
\includegraphics[trim = 7mm 0mm 0mm 0mm, clip,height=6.8cm,width=\columnwidth]{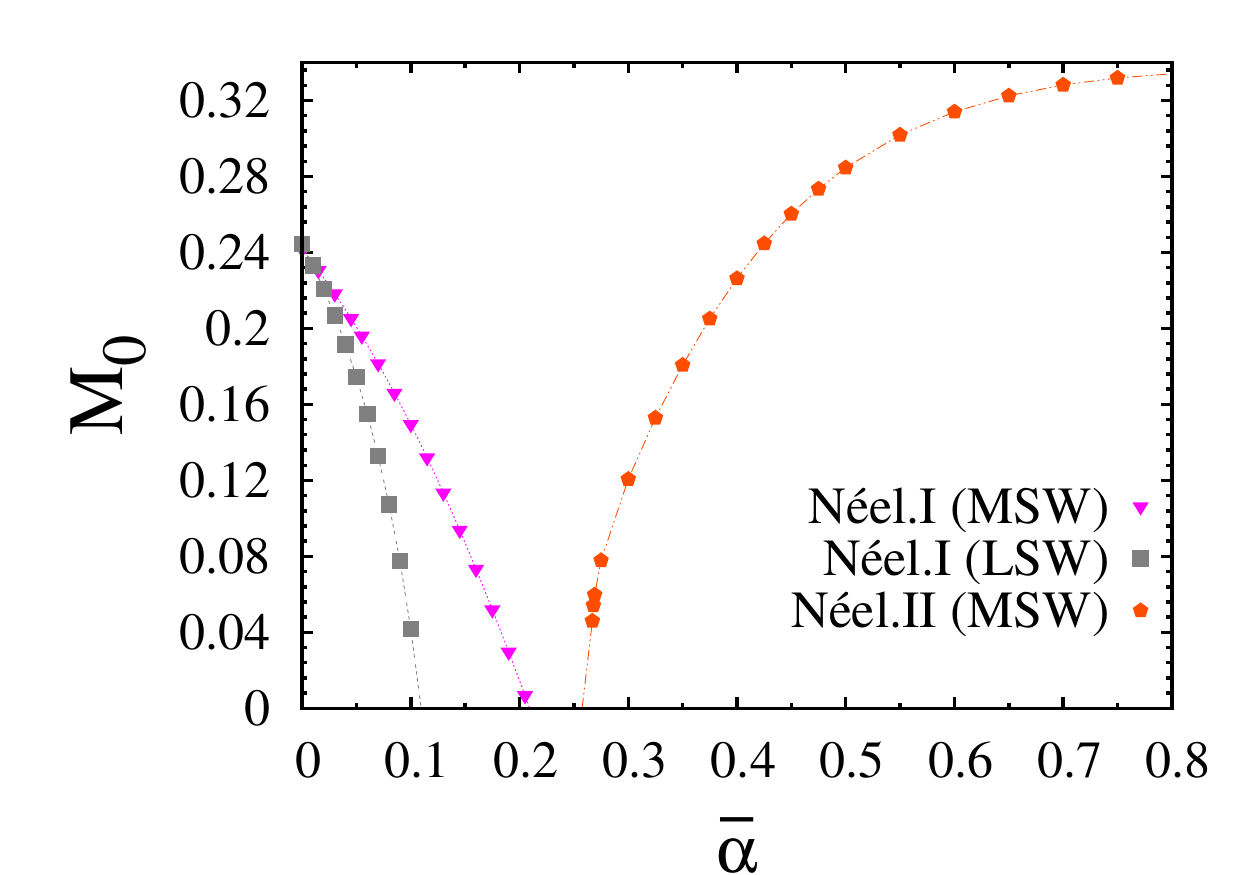} 
\caption{(Color online) Spontaneous magnetization ($M_0$) per site of  
symmetric $J_1-J_2$ model versus the frustration parameter ${\bar\alpha}$ for $S=1/2$.  The pink triangles and red diamonds  show the magnetization obtained from MSW in N{\'e}el.I and N{\'e}el.II phases, respectively. The grey squares represent the magnetization in  LSW approximation.} 
\label{magnetic} 
\end{figure} 
\begin{figure}[b] 
\centering 
\includegraphics[trim = 7mm 0mm 0mm 0mm, clip,width=\columnwidth]{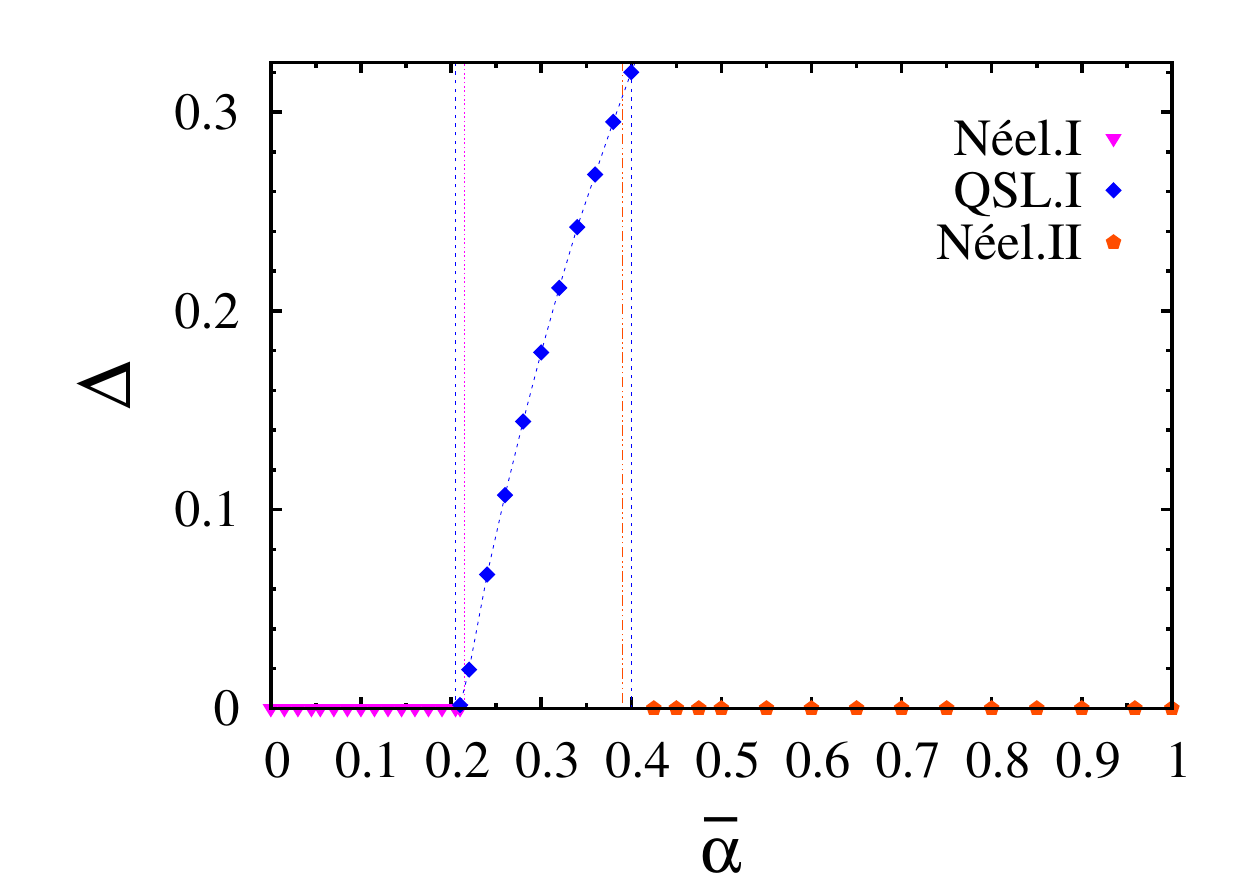} 
\caption{(Color online) Energy gap of magnon spectrum ($\Delta$) (in units of $J_1$) versus  the frustration parameter ${\bar\alpha}$. $\Delta$ is zero in the 
N{\'e}el.I ordered phase and then grows continuously in QSL.I phase and finally drops to zero in the N{\'e}el.II ordered phase.    
} 
\label{gap} 
\end{figure} 
 \begin{figure} 
  
\begin{subfigure}[~N{\'e}el.I phase]{\includegraphics[width=\columnwidth]{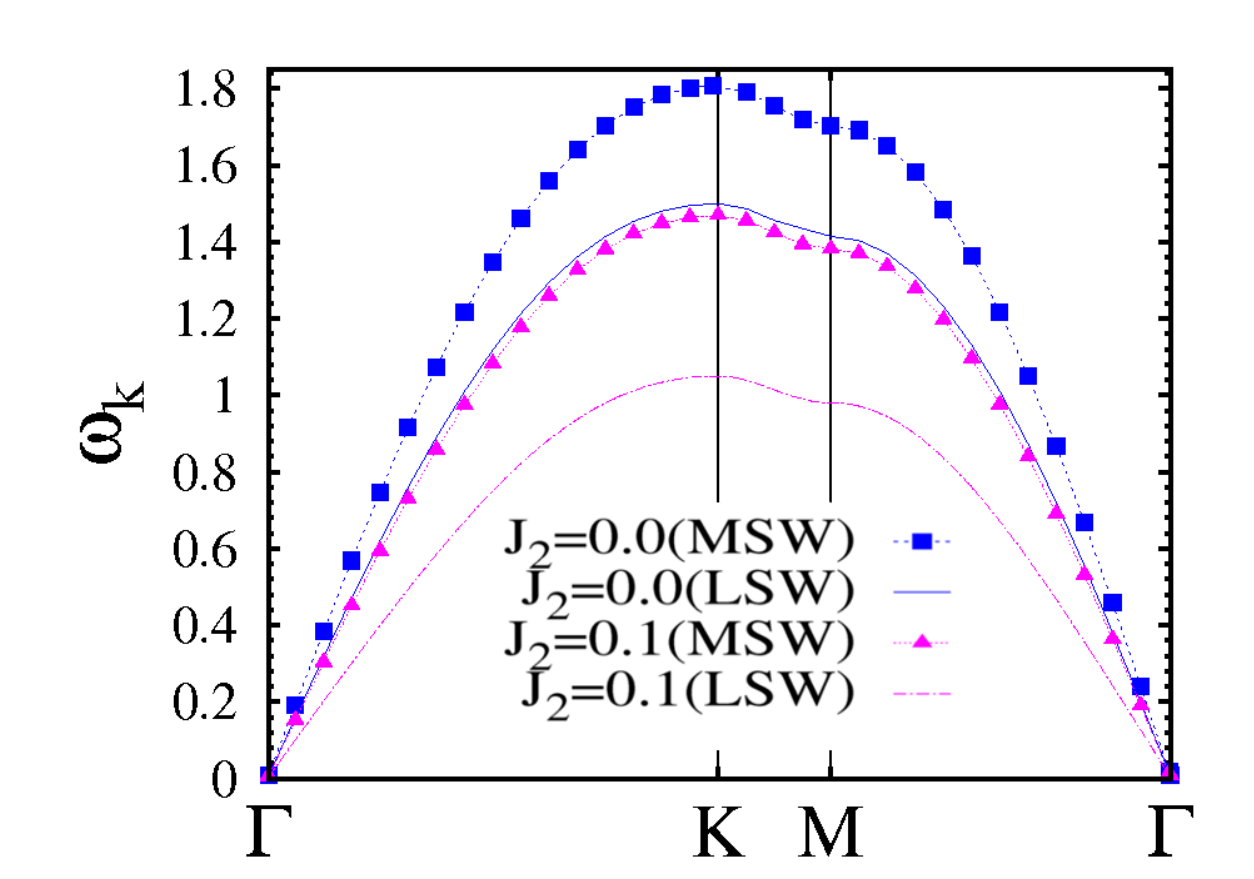}} 
\end{subfigure} 
  
\begin{subfigure}[~QSL phase]{\includegraphics[width=\columnwidth]{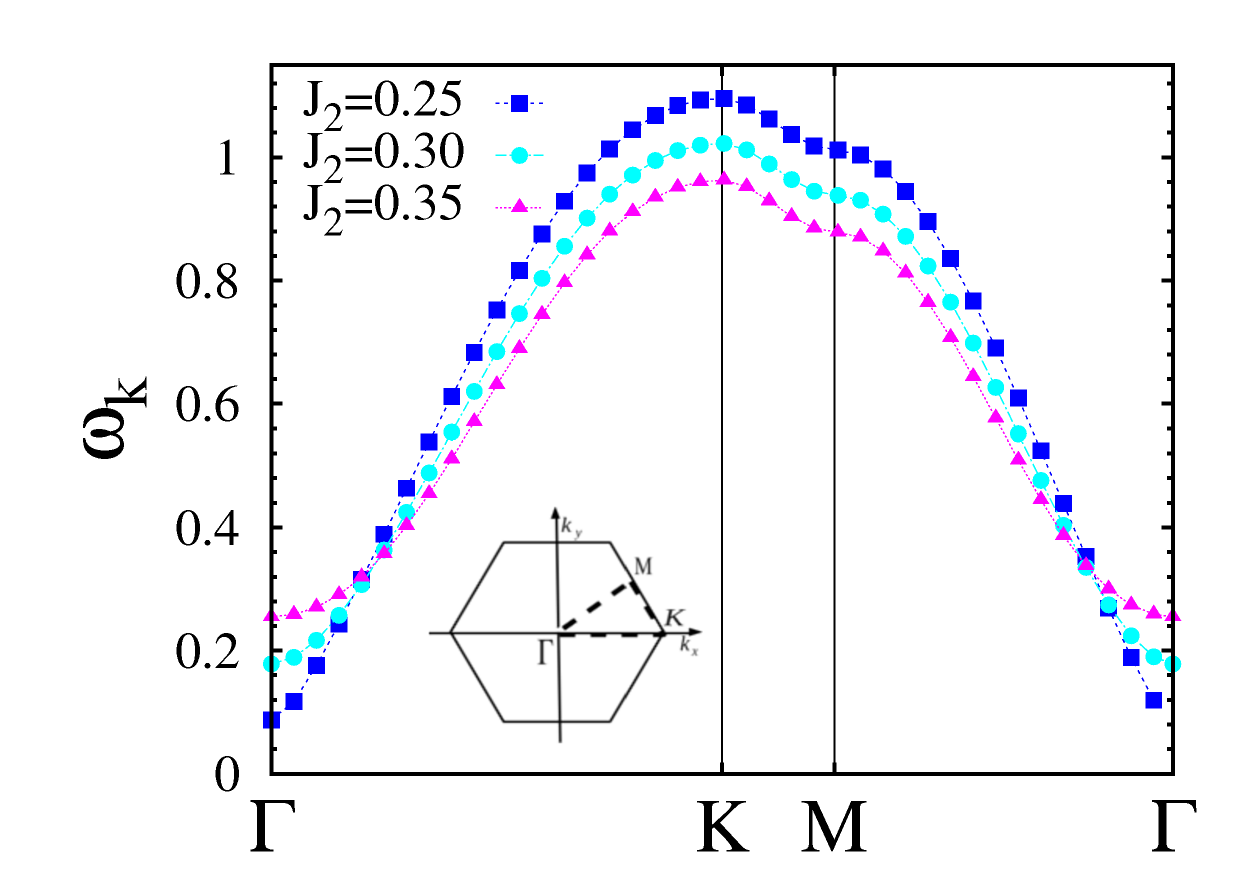}} 
\end{subfigure} 
  
\begin{subfigure}[~N{\'e}el.II phase]{\includegraphics[width=\columnwidth]{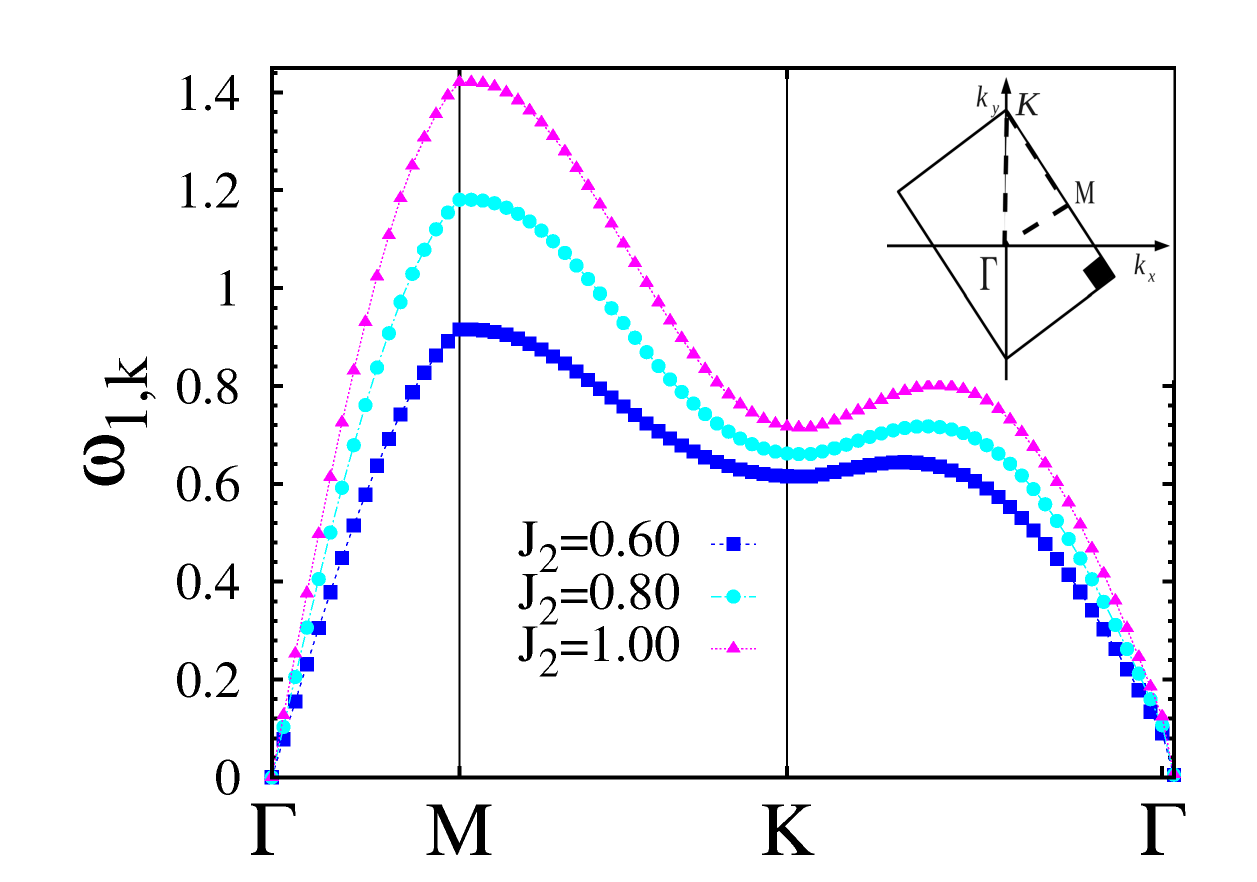}} 
\end{subfigure} 
  
  
\caption{(Color online) Magnon energy  dispersion $\omega_k$ (in units of $J_1$) along symmetry directions in the magnetic first Brillouin zone (1BZ)  for  (a) N{\'e}el.I, (b) QSL, and  (c) N{\'e}el.II phases. The insets illustrate the magnetic 1BZ corresponding to each phase. The magnetic 1BZ of N{\'e}el.I state is the same as the 1BZ of the honeycomb lattice as shown by the inset of panel (b). The solid and dashed lines in the panel (a) represent the magnon dispersion obtained in LSW approximation.} 
 \label{omega} 
\end{figure} 
 \begin{figure}[t] 
\centering  
{\includegraphics[width=\columnwidth]{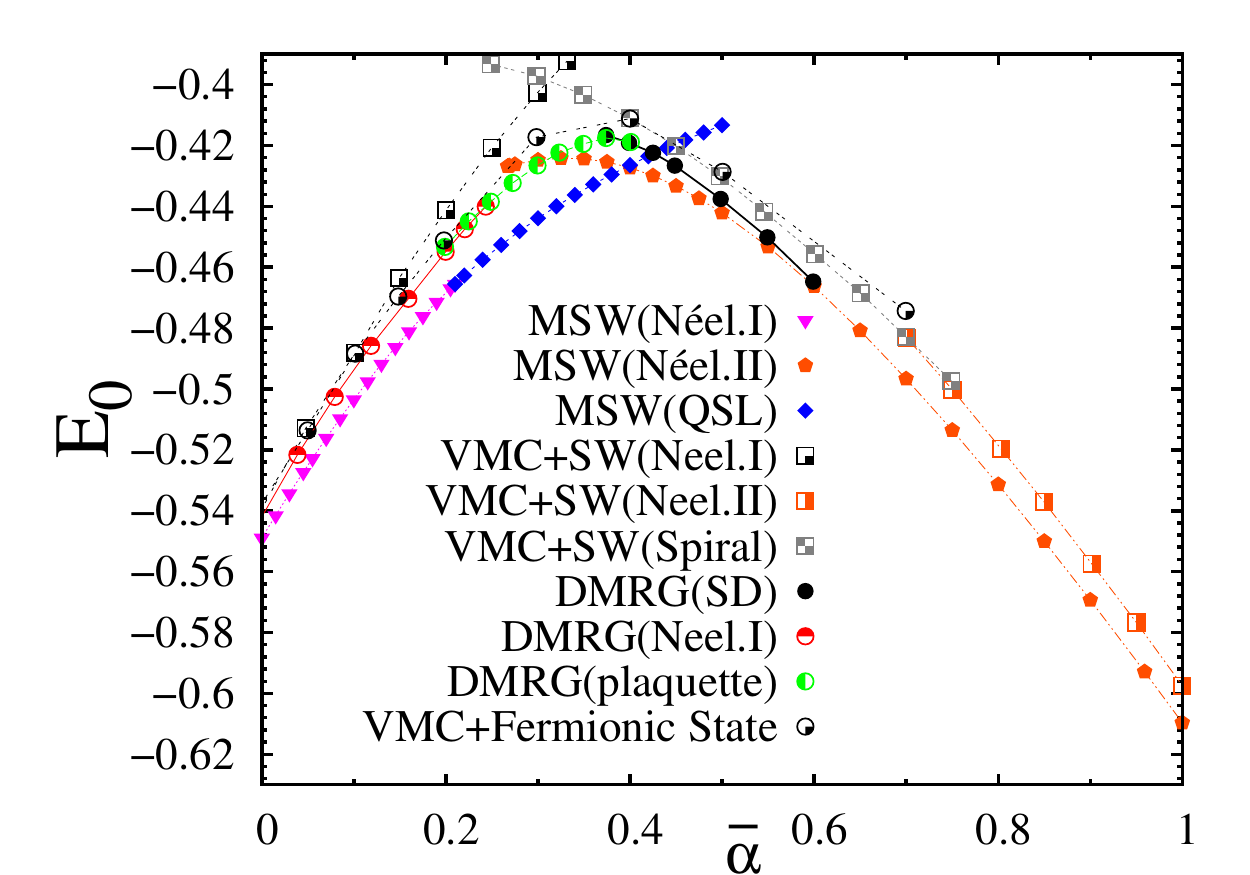}} 
{\includegraphics[width=\columnwidth]{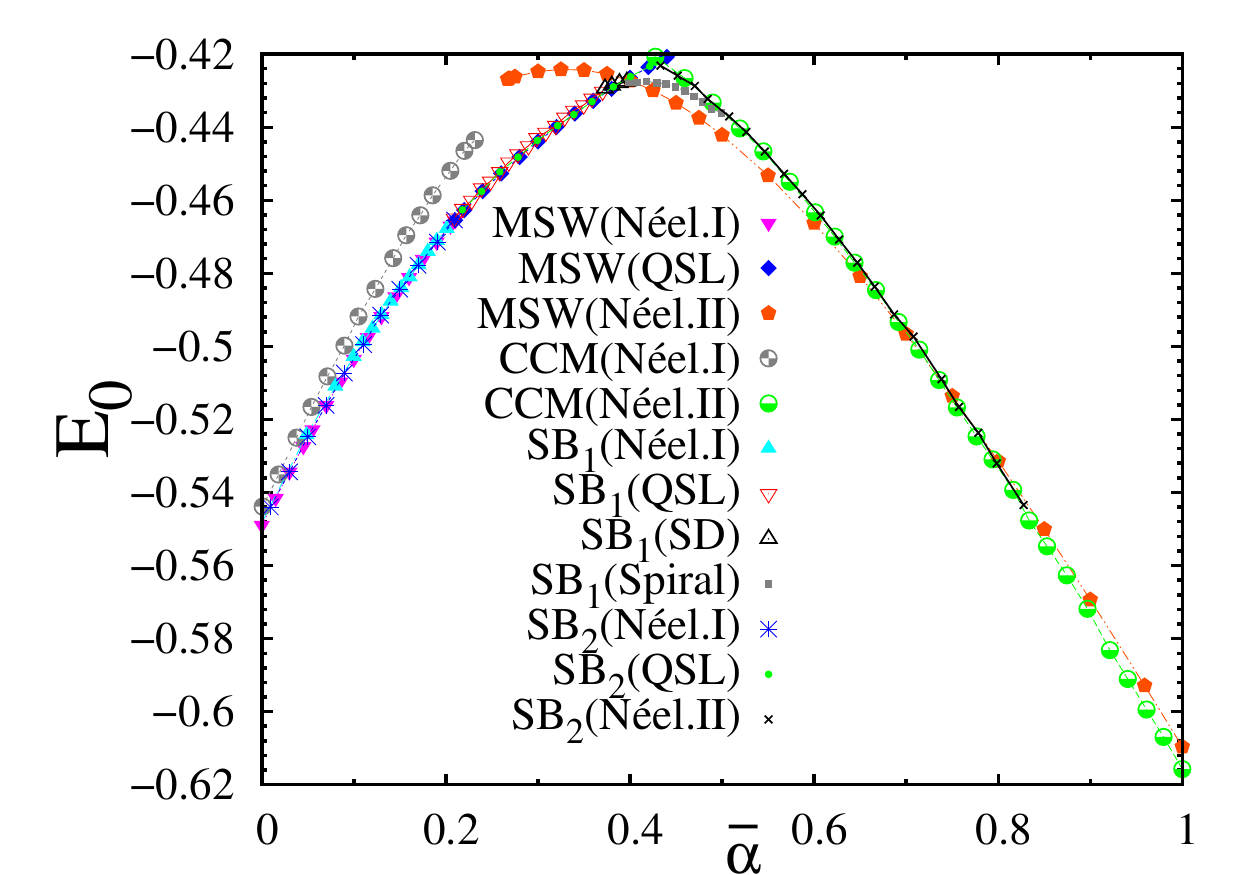}} 
\caption{(Color online) Comparison of MSW results with ({\bf {top}}) density-matrix renormalization group (DMRG)\cite{pvb7}, variational Monte Carlo (VMC) based on Jastrow and projected fermionic states (VMC+fermionic state) \cite{VMC-sondhi} and spin wave states (VMC+SW state) \cite{Ciolo}, and ({\bf {bottom}}) the coupled cluster method (CCM) \cite{pvb4} and Schwinger boson  approach (SB$_1$~\cite{SB-lamas1} and SB$_2$~\cite{SB-china}. SD denotes  staggered dimerized state.
} 
\label{compareE} 
\end{figure} 

  

Having a  long range ordered (LRO) ground state, requires  the gapless excitation spectrum  as  the result of Goldstone theorem. This condition 
leads to vanishing of  the chemical potential $\mu$ (defined by Eq. \eqref{minn}), in the ordered state  as a requirement of BEC transition~\cite{cirac}. 
To calculate the energy and magnetization for each type of ordering one needs to solve the self-consistent equations \eqref{f}, \eqref{g}, \eqref{M}, \eqref{A1} and \eqref{B1}, with $\mu=0$.  
After convergence, these  equations give  the spontaneous magnetization $M_0$ and the functions  $f_{ij}$  and $g_{ij}$, then substitution of $f_{ij}$ and $g_{ij}$ in equation \eqref{finalH} gives the ground state energy per site $E_0$.  The magnon excitation spectrum is given by equation \eqref{magnon1}, and spin-spin correlations can be calculated by the equations \eqref{ss-correlation1} and \eqref{ss-correlation2}.   
For N{\'e}el.II and III orderings, it is more convenient to use a four-sublattice unite cell (Fig.\ref{ordered}-b,c), wherefore   the  ordering wave vector is $Q=0$. Using the larger  unit cell in real space leads to reduction of the size of magnetic Brillouin zone  in ${\bf K}$-space and so the number of singular points, hence making   the convergence of corresponding self-consistent equations much easier (see Appendix \ref{app.2} for details). 
In this case,  the physical quantities of interest can be calculated by solving the set of equations \eqref{fg2}.  

Following the above procedure, MSW results that within  the  possible ordered state, The N{\'e}el.I and N{\'e}el.II acquire the minimum energy for  $0\leq {\bar\alpha}\lesssim 0.207$ and  ${\bar\alpha}\gtrsim 0.25$, respectively.  
  The dependence of  ground state energy per site ($E_0$) and corresponding spontaneous magnetization ($M_0$) on the frustration parameter ${\bar\alpha}$ are illustrated in figures \ref{energy} and \ref{magnetic}, respectively.
In figures \ref{energy} and \ref{magnetic}, $E_0$ and $M_0$ obtained by the linear spin wave (LSW) theory,  are also represented for N{\'e}el.I state. LSW indicates that the  N{\'e}el.I phase is stable only to ${\bar\alpha}\approx 0.11$. Therefore, a  comparison between LSW and MSW shows that the the nonlinear interactions, taken into account  by MSW in mean field approximation,  lower the energy of the N{\'e}el.I phase  and also increases its stability against the frustration  up to ${\bar\alpha}\approx 0.207$.  

On the other hand, for ${\bar\alpha> 0.25}$ we found that  N{\'e}el.II has lower energy with respect to the N{\'e}el.III and also the spiral states.
 The energy per site and magnetization, corresponding to this type of ordering, is plotted for the range ${0.25<\bar\alpha\leq 0.8}$ in figures \ref{energy} and \ref{magnetic}. It is important to mention that N{\'e}el.II state is not a classically stable state. Indeed assuming such an ordering and using  LSW approximation, it is found that complex numbers appear in its spin excitation spectrum which makes this state unstable. Hence, the stability of this phase in MSW can be attributed to the nonlinear magnon-magnon interactions. 

For the interval ${0.207 \lesssim \bar\alpha\lesssim 0.25}$, however, no ordered state is found to be stable. Indeed, the magnetization of N{\'e}el.I state falls continuously to zero at ${\bar\alpha\approx 0.207}$, above which no stable solution of self-consistent equations corresponding to N{\'e}el.I ordering is possible with the BEC condition $\mu=0$. However, starting from N{\'e}el.I state and relaxing the BEC condition and setting $M_0=0$, it is possible to obtain from MSW equations a magnetically disordered state with finite chemical potential and vanishing magnetization for ${\bar\alpha}\gtrsim 0.207$.  In this case  the chemical  potential, $\mu$, has to be considered as a quantity which is to be found self-consistently. 

In addition to the $SU(2)$ symmetry of the spin Hamiltonian \eqref{j1j2-distort}, such a  disordered phase preserves  all the symmetries of the lattice, i.e. the $C_3$ and $C_6$ rotational and translational symmetries. In fact all the attempts to find  a solution with broken rotational symmetry, for example a solution with not equal pairing and hopping functions on different bonds, were unsuccessful. Such a magnetically disordered state which respects all the symmetries of the Hamiltonian and the lattice is called quantum spin liquid (QSL)  state. As it can be seen from the figure \ref{energy}, the energy curve of the N{\'e}el.I state connects smoothly to the QSL state, an indication of a continuous phase transition between these two ground states. Moreover the calculation of spin gap, illustrated in figure \ref{gap}, shows  the continuous rise of the magnon gap in this phase.  Interestingly, the stability of QSL state goes beyond ${\bar\alpha=0.25}$ and its energy is lower than the N{\'e}el.II phase up to ${\bar\alpha}\approx 0.397$ where  it crosses the energy curve of  N{\'e}el.II. As a conclusion, the transition between these two phases are first order. Figure \ref{gap} shows that at this  transition point  the spin gap drops discontinuously to zero. 

Since these gapped QSL phase is obtained by starting from the  N{\'e}el.I state it possess all the symmetries of N{\'e}el.I, hence we call it QSL.I. Starting from N{\'e}el.II and III, it is also possible to find  QSL states, however with higher energy with respect to  QSL.I. Calculation of spin-spin correlations for  QSL.I shows the existence short-range N{\'e}el.I type correlations in this phase (see Table.\ref{tab-correlation} and figure \ref{phase-corr}-b).   

Figure \ref{omega}, represents the magnon dispersion along the symmetry directions in the magnetic Brillouin Zone of for  N{\'e}el.I (panel-(a)), QSL (panel-(b)) and N{\'e}el.II (panel-(c)) phases. In panel-(a), the LSW magnon dispersion is also shown to have  lower energy with respect to the MSW dispersion,  indicating the more rigidity of ordered phase as a result of magnon-magnon interaction.  In panel-(c) of figure \ref{omega} only the lower branch  of magnon dispersion, given by equation \eqref{magnon2}, is plotted.

To close this section, we compare the MSW results  with some other methods. Figure \ref{compareE} displays such a comparison, in top panel of this figure the MSW ground state energies are co-platted with similar results obtained by      
density-matrix renormalization group (DMRG)~\cite{pvb7}, variational Monte Carlo (VMC) approaches based on  projected fermionic states (VMC+fermionic state)~\cite{VMC-sondhi}, VMC based on  spin wave states (VMC+SW state)~\cite{Ciolo}.
The comparison of MSW results with  CCM~\cite{pvb4} and Schwinger boson (SB) mean field  approaches 
SB$_1$~\cite{SB-lamas1} and SB$_2$\cite{SB-china} is also illustrated  in the bottom panel of this figure. 

The top panel  clearly shows that the MSW ground state energies in the three phases lay below the energies obtained by DMRG and VMC. Specifically,  for the disordered region the QSL state proposed by MSW has lower energy with respect to the plaquette valence bond (PVB)  and also staggered dimerized (SD)  (both proposed by DMRG).  For ${\bar\alpha\gtrsim 0.4}$, the N{\'e}el.II state obtained by MSW has lower energy than the spiral state proposed by VMC+SW~\cite{Ciolo}. VMC base upon fermionic states yields a N{\'e}el.I phase for $0\leq{\bar\alpha}\lesssim 0.08$, a $Z_2$ QSL state for $0.08\lesssim{\bar\alpha}\lesssim0.3$ and a SD state for ${\bar\alpha}\gtrsim 0.3$~\cite{VMC-sondhi}

 On the other hand, The bottom panel shows that MSW results are in a very good agreement  with Schwinger boson (SB) mean filed approach~\cite{SB-lamas1,SB-china}  in the N{\'e}el.I and QSL phases, while it gives lower energy for this phase with respect to CCM~\cite{pvb4}. For the N{\'e}el.II state, although the MSW result  agrees well with SB$_2$ and CCM for ${\bar\alpha}\gtrsim 0.6$, nevertheless its energy lays below the ones obtained by other two for ${\bar\alpha}\lesssim 0.6$.  Hence, the transition point from QSL to N{\'e}el.II which is  ${\bar\alpha\approx 0.42}$ for SB$_2$, moved to a smaller  value ${\bar\alpha\approx 0.396}$ for MSW.  Moreover, by calculation of PVB and SD susceptibilities, CCM predicts a PVB state for $0.207<{\bar\alpha}<0.385$ and SD ground states  for $0.385<{\bar\alpha}<0.65$. 
 SB$_1$~\cite{SB-lamas1} also results in a SD ordering for $0.373\lesssim{\bar\alpha}\lesssim 0.398$ with a competitive   energy  with the  QSL.I found by MSW. SB$_1$ also gives rise to a spiral ground state for $0.398<{\bar\alpha}<0.5$ with a larger energy than the N{\'e}el.II obtained by MSW.  
 
Like MSW,  the entangled pair variational ansatz (not shown in Fig.\ref{compareE})~\cite{eps} yields  a N{\'e}el.I ordered state  for $0\leq {\bar\alpha} \lesssim 0.2$, a N{\'e}el.II state  for $0.4\lesssim{\bar\alpha} <1$ and a symmetry preserving disordered state for $0.2\lesssim {\bar\alpha} \lesssim 0.4$, though with higher energies respect to MSW.

\section{phase diagram of distorted model}
\label{j1j2-dis}

\begin{figure}[t]
\includegraphics[trim = 6mm 0mm 0mm 0mm, clip, width=\columnwidth]{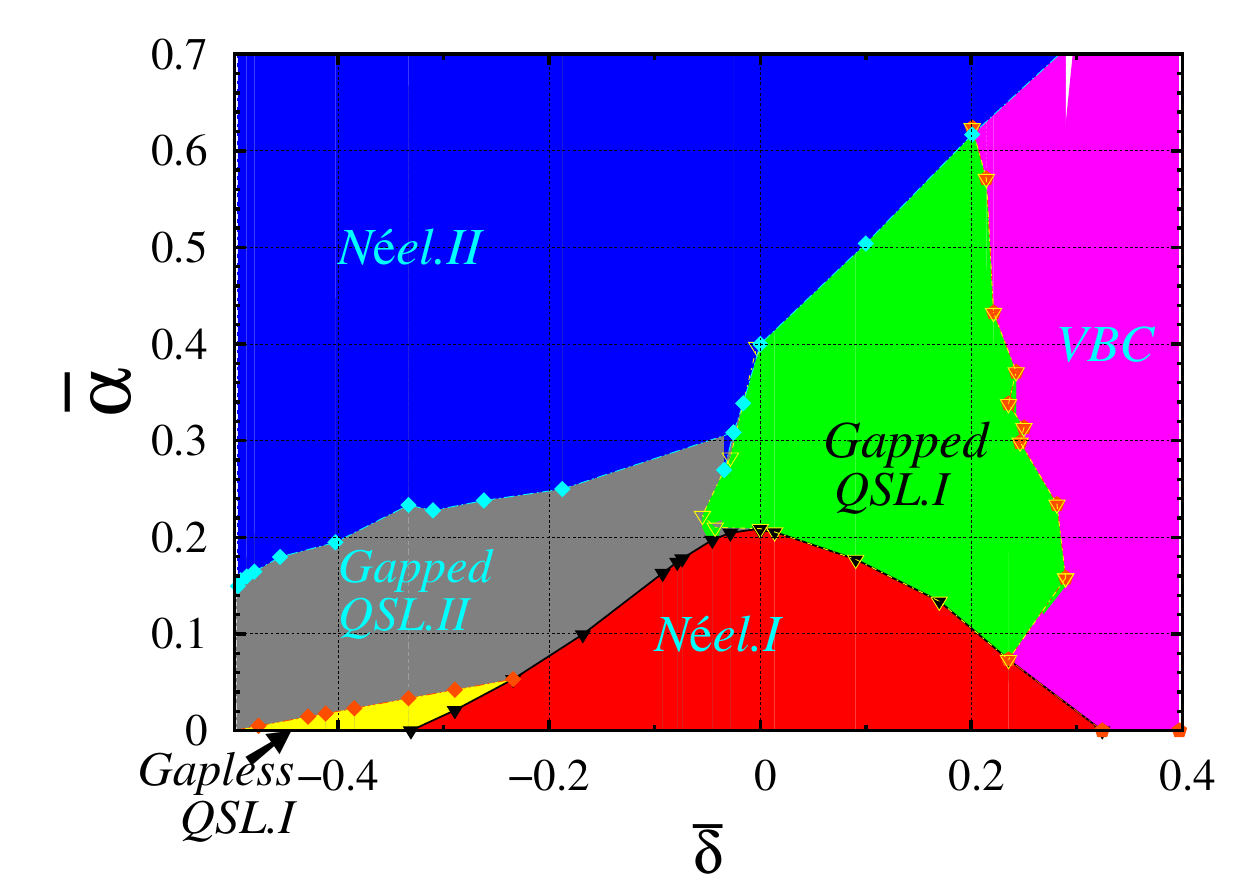} 
\caption{(Color online) MSW phase diagram of $S=1/2$ distorted $J_1-J_2$ honeycomb anti-ferromagnet. QSL.I and II denotes the quantum spin liquid states originating from N{\'e}el.I and N{\'e}el.II, respectively. VBC stands for valence bond crystal state.   
 } 
\label{distorted-phase} 
\end{figure} 

In this section we discuss the phase diagram of  $S=1/2$ distorted model Hamiltonian \eqref{j1j2-distort}. The MSW phase diagram of the model is represented in figure \ref{distorted-phase} in plane  of the distortion parameter ${\bar\delta}$ and frustration ${\bar\alpha}$. 
Like the symmetric model, magnetically ordered phase in the presence of distortion are found to be the collinear states N{\'e}el.I and II.

Figure \ref{distorted-phase}, shows that the maximum stability of N{\'e}el.I state occurs for isotropic model $\bar\delta=0$. Distortion  in both c positive  ($\bar\delta >0$ i.e. $J_1>J'_1$) and  negative  ($\bar\delta<0$ i.e. $J_1<J'_1$) cases,  makes this phase more fragile  against frustration.  For $|\bar\delta|\approx 0.3$, N{\'e}el.I phase becomes totally unstable for any $\bar\alpha>0$. The stability region  of N{\'e}el.I state versus distortion is in agreement with the results of renormalization group (RG) calculations done  on the nonlinear sigma model (NLSM) presentation of the model~\cite{takano}.  However, the RG-NSLM underestimates the stability range of this phase against frustration, i.e.  finds the maximum stability range  $0\leq{\bar\alpha\lesssim 0.11}$ for the symmetric model (${\bar\delta=0}$).  

For N{\'e}el.II phase, while the positive distortion ($\bar\alpha>0$) has a destructive effect on the stability of this phase against the frustration, nevertheless, negative distortion ($\bar\delta<0$) extends its stability to lower value of frustration.  Note that in order to make  $J_1$ and $J'_1$ positive, the distortion parameter should be in the interval $[-0.5,1.0]$.

In addition to these to ordered  disordered phases we find four magnetically distinct disordered phases, (i) a valence bond crystal (VBC) phase for large positive distortion, (ii)  a gapped QSL  originating form the N{\'e}el.I state (gapped QSL.I) for intermediate positive and small negative distortions, (iii)  a gapped QSL originating from the N{\'e}el.II state(gapped QSL.II) for negative distortions and intermediate frustration and (iv)  gapless QSL originating from N{\'e}el.I  (gapless QSL.I) for large negative distortion and small frustration. Apart from the gapped QSl.II, all the other three disordered phase VBC, gapped and gapless QSL.I are the self-consistent solutions of MSW equations  with started from the N{\'e}el.I ordering state, but with vanishing spontaneous  magnetization. On the other hand, in the stability region of gapped QSl.II, starting from N{\'e}el.I ordering, the self-consistent equations does not converge to any stable solution. However, in this region assuming a N{\'e}el.II type ordering, a stable disordered state comes out of MSW equations.

\begin{figure*}
\includegraphics[width=\textwidth]{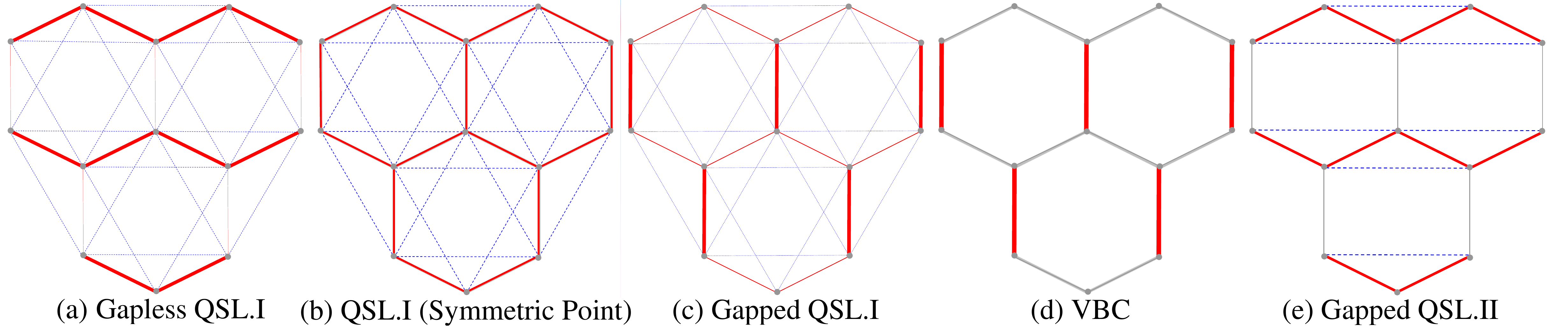}
\caption{(Color online) Schematic representation of spin-spin correlations for the first and second neighbors for a representative pint in the phases: (a) gapless QSL.I, (b) symmetric gapped QSL.I ($\bar\delta=0$), (c) distorted gapped QSL. I, (e) valence bond solid (VBC) and (d) gapped QSL.II. The solid red and dotted blue lines denote  negative  (AF) and positive (F) correlations. The thickness of the lines are proportional to the correlation magnitude.}
\label{phase-corr}
\end{figure*}

 To gain insight into nature of the disordered states, we calculated the spin-spin correlation for the nearest and next to nearest neighbor spins.  The correlation data are given in Table.\ref{tab-correlation} for a representative point in each phase. These results are also displayed schematically  in figure \ref{phase-corr}.  

As it is clear from the first three rows of Table.\ref{tab-correlation} and panels (a),(b) and (c) of figure \ref{phase-corr}, in QSL.I there are short-range correlation inherited  from  N{\'e}el.I ordering, i.e. nearest neighbor negative (AF) and next to nearest neighbor positive (F) correlations. In absence of distortion ($\bar\delta=0$) the correlations are the same in all directions. However, in the presence of distortion, the AF correlations are stronger for the nearest neighbor bonds with larger exchange coupling (figure~\ref{phase-corr}-(a) and (c)).

 While for positive and small negative distortion the QSL.I state is gapped, for small frustration and large negative distortions (say $-0.5 \leq \bar\delta\lesssim -0.3$) this phase is gapless.  
It can be seen from the first row of Table.\ref{tab-correlation} that the AF correlations along $J'$-bonds are larger than the one along $J$-bond, by two orders of magnitude, hence if $J_2$ is small enough, the honeycomb spin system in this case can be considered as a system of weakly coupled chains with coupling $J'$. Therefore,  the fact that ground state of a $S=1/2$ Heisenberg chain is a gapless  spin liquid state, would  be a justification for QSL.I state in this region being gapless. 

For large positive distortions, there are vanishing correlations between  the nearest neighbor correlations in $\delta_2$ and $\delta_3$ directions as well as between all the second neighbors. In this case, the spins residing on $J_1$-bonds ($\delta_1$ directions) form strong singlet valence bonds (figure \ref{phase-corr}-(d)). In such a strong dimerized state,  singlets are prevented from hopping  to the neighboring  bonds and so are frozen. This is the reason for a calling it a valence bond crystal (VBC).  

Finally, in the region of the stability for gapped QSL.II (negative distortions and moderate frustration)  the AF correlation along $J_1$-bonds as well as the positive correlations along $\delta_1$ and $\delta_2$ correlations  are negligible (the last row of Table.\ref{tab-correlation} and figure \ref{phase-corr}-(e)).  In this phase the system can also be considered as  effectively decoupled chains with nearest negative and next to nearest positive correlations. It seems the enhanced frustrating interaction $J_2$ between the second neighbors pushes the two spins within each unit cell into their high spin state, hence, roughly speaking, this spin system can be effectively  described by an $S=1$ chains for which spin excitations are gapped.

\begin{table*}[] 
\caption{Spin-spin correlation functions of nearest neighbors ( $\langle S_{\bf{i}}\cdot S_{{\bf i}+{\bf\delta}_1}\rangle $, $\langle S_{\bf i}\cdot S_{{\bf i}+{\bf\delta}_2}\rangle $, $\langle S_{\bf{i}}\cdot S_{{\bf i}+{\bf\delta}_3} \rangle $) and next to nearest neighbors  ($\langle S_{\bf{i}}\cdot S_{{\bf i}+{\bf\delta}^\prime_1}\rangle, \langle S_{\bf i}\cdot S_{{\bf i}+{\bf\delta}^\prime_2}\rangle, \langle S_{\bf{i}}\cdot S_{{\bf i}+{\bf\delta}^\prime_3}\rangle$) in different phases of distorted honeycomb anti-ferromagnet. The vectors ${\bf\delta}_1, {\bf\delta}_2, {\bf\delta}_3, {\bf\delta}^\prime_1, {\bf\delta}^\prime_2$ and ${\bf\delta}^\prime_3$ are shown in Fig. \ref{honeycomb}.} 
\centering 
\begin{tabular}{|c| c |c |c |c |c |c|} 
\hline 
${\bar\alpha}$ & ${\bar\delta}$ & $\langle S_{\bf{i}}\cdot S_{{\bf i}+{\bf\delta}_1}\rangle $ & $\langle S_{\bf{i}}\cdot S_{{\bf i}+\delta_{2,3}}\rangle$ & $\langle S_{\bf{i}}\cdot S_{{\bf i}+\delta^\prime_{1,2}}\rangle $ & $\langle S_{\bf{i}}\cdot S_{{\bf i}+\delta^\prime_{3}}\rangle $ & State\\
 \hline 
$0.02$ &$-0.4$        & $-0.0065$     &  $-0.3276$  &  $0.0282$   &  $0.0282$     & Gapless QSL.I \\[1ex]
$0.25$ & $0$       & $-0.1720$    &    $-0.1720$   &   $0.0411$ &   $0.0411$  &  gapped QSL.I (symmetric) \\  [1ex]  
$0.3125$ & $0.125$     & $-0.3217$    & $-0.05680$    &  $0.0122$ &   $0.0122$  &    Gapped QSL.I\\ [1ex]   
$0.425$ & $0.350$     & $-0.3753$     &  $-4.2\times10^{-7}$    &  $5.0\times10^{-8}$ & $5.0\times10^{-8}$ &VBC\\ [1ex]
$0.1875$ & $-0.125$    & $3\times 10^{-11}$ &    $-0.2230$    & $-4.3\times10^{-9}$ &   $0.0615$ &  Gapped QSL.II\\  [1ex] 
\hline 
\end{tabular} 
\label{tab-correlation} 
\end{table*}


 \section{conclusion}
\label{conclusion} 

Taking advantage of DM transformation, which are exact and hence unlike Holstein-Primakoff transformation need not be truncated,  
 MSW provides  a powerful tool to extract the phase diagram of spin systems. Using MSW, we explored the ground state of symmetric and distorted $S=1/2$  Heisenberg $J_1-J_2$ antiferromagnet in honeycomb lattice. For the symmetric model, where all equivalent bonds in honeycomb lattice have equal exchange couplings, we found two types of collinear ordering  in small and large  frustration limit, namely a two-sublattice ordering N{\'e}el.I for $0\leq J_2/J_1 \lesssim 0.207$ and a four-sublattice ordering N{\'e}el.II  for $J_2/J_1\gtrsim 0.396$. The N{\'e}el.II is not a classical solution and so is unstable when quantum fluctuations are taken into account by linear spin wave theory.  Indeed, for $S=1/2$ the enhanced nonlinear quantum fluctuations  tend to stabilize this phase.  
 For intermediate frustration $0.207 \lesssim J2/J1 \lesssim 0.396$ a magnetically disordered state which preserves all the symmetries of the system is found to be the ground state that is a gapped  QSL.
The short-range correlations in this QSL has the symmetries  of N{\'e}el.I, then we coined the name QSL.I for this phase.
We found that these two phases transform to each other by a  continuous phase transition. However, symmetries of the QSL.I are different from N{\'e}el.II, and so a first order transition is found between these two states as expected.  As a conclusion the order-disorder transitions in this system can be described in the framework  of Landau-Ginzburg theory.   

Introducing the distortion  to the model  breaks its $C_3$ symmetry. This leads to the emergence of new phases as the result of  the interplay between distortion and frustration. These new phases, all being magnetically disordered, are a gapless QSL.I originating from  N{\'e}el.I ordering, a gapped QSL.II originating from  N{\'e}el.II and a valence bond solid state where the singlet dimers are frozen on the bonds with larger coupling. We discussed that in both gapless QSL.I and gapped QSL.II phases, the model can be effectively be described  in terms of weakly coupled zigzag chains. 

The main privilege  of MSW over other methods, such as DMRG, VMC and ED,  is that it is free from finite size effect. However, validity of  the mean-filed approximation incorporated  in this method might be under  question  when the quantum fluctuations become large. The quantum fluctuations are significantly large in the disordered states where  the spontaneous magnetization, or in terms of bosons the condensate,  vanishes. This suggest that the QSL states proposed for the disordered region of the phase diagram has to be considered cautiously.   Therefore, to improve the validity of MSW states,  one could consider them as the initial wave function for the variational methods.



\appendix 
  
\section{Derivation of  MSW self-Consistent equations  } 
\label{app.1}
To diagonalize the Hamiltonian \eqref{DMH} in mean field approximation, we need to define the  Bogoliubov transformations 
\begin{eqnarray} 
\alpha_{\bf k}&=&\cosh(\theta_{\bf k})a_{\bf k}-\sinh(\theta_{\bf k})b_{-{\bf k}}^\dag,\nonumber\\ 
\beta_{-\bf k}^\dag&=&-\sinh(\theta_{\bf k})a_{\bf k}+\cosh(\theta_{\bf k})b_{-{\bf k}}^\dag,\nonumber\\ 
\label{BogoN} 
\end{eqnarray} 
  where $a_{\bf k}$ and $b_{-\bf k}^\dag$ are the Fourier transformations of $a_{i}$ and $b_{j}^\dag$ (defined by equation \eqref{DM}), 
\begin{eqnarray} 
a_{\bf k}&=&\sqrt{\frac{2}{N}}\sum_{i\in A}e^{-\bf{ik}\cdot{\bf r}_i}a_{i},\nonumber\\ 
b_{-\bf k}^\dag&=&\sqrt{\frac{2}{N}}\sum_{j\in B}e^{-\bf{ik}\cdot{\bf r}_j}b_{j}^\dag,\nonumber\\ 
\label{Ft} 
\end{eqnarray} 
in which $N$ is the total number of sites. The mean field Hamiltonian in its  diagonalized form, in terms of noninteracting  Bogoliubov quasiparticles, is written as   
\begin{equation} 
\begin{split} 
&H_{\rm MF}=\sum_{\bf k}[\omega_{\bf k}(\alpha_{\bf k}^\dag\alpha_{\bf k}+\beta_{\bf k}^\dag\beta_{\bf k})]+NE_0, 
\end{split} 
\label{E} 
\end{equation} 
where $\omega_{k}$ is the excitation energy spectrum and $E_0$ is the ground state energy per site given by equation \eqref{Ehoney}. Substituting $a_i$ and $b_i$ in equation \eqref{relation} in terms of   Bogolon operators   \eqref{BogoN}, for a pair of DM bosons at a given displacement vector  ${\bf r}_{ij}={\bf r}_{i}-{\bf r}_{j}$,  one finds for hopping ( $f_{ij}=f({\bf r}_{ij})$) and pairing ($g_{ij}=g({\bf r}_{ij})$)  expectation functions
\begin{equation} 
f_{ij}=\frac{1}{N}\sum_{\bf k}^\prime\cosh(2\theta_{\bf k})\exp(-{\bf ik}\cdot{\bf r}_{ij}), 
\label{f1} 
\end{equation} 
with $\ i,j \in A$  or $B$, and 
\begin{equation} 
g_{ij}=\frac{1}{N}\sum_{\bf k}^\prime\sinh(2\theta_{\bf k})\exp(-{\bf ik}\cdot{\bf r}_{ij}),
\label{g1} 
\end{equation} 
with $i\in A$ and  $j\in B$. Otherwise $f_{ij}$ and $g_{ij}$  vanish. 
In equation \eqref{f1} and \eqref{g1}, $\sum_{\bf k}^\prime$ denotes  the sum of  over half  of the Brillouin zone. 

We then minimize the mean filed energy \eqref{finalH} with respect to $\theta_{k}$, under the constraint \eqref{constraint}, that is 
\begin{equation} 
\frac{\partial[E-\mu f(0)]}{\partial \theta_{\bf k}}=0,\\ 
\label{minn} 
\end{equation} 
  where  the Lagrange multiplier $\mu$ can considered the  chemical potential  needed to fix the number of DM bodons in order to fulfill the Takahashi's   constraint.   Minimization \eqref{minn} yields the following  set of self-consistent equations  
\begin{equation} 
f_{ij}=M_0+\frac{1}{N}\sum_{{\bf k}\neq {\bf 0}}^\prime\frac{ B_{\bf k}}{\omega_{\bf k}}e^{\bf{ik}\cdot\bf{r}_{ij}}, 
\label{f} 
\end{equation} 
\begin{equation} 
g_{ij}=M_0+\frac{1}{N}\sum_{{\bf k}\neq {\bf 0}}^\prime\frac{ A_{\bf k}}{\omega_{\bf k}}e^{\bf{ik}\cdot\bf{r}_{ij}},
\label{g} 
\end{equation} 
and
\begin{equation} 
M_0=S+\frac{1}{2}-\frac{1}{N}\sum_{\bf k\neq 0}^\prime \frac{B_{\bf k}}{\omega_{\bf k}}. 
\label{M} 
\end{equation} 

here  $A_{\bf k}$ and $B_{\bf k}$ are given by
\begin{equation} 
A_{\bf k}=\frac{1}{2}\sum_{\delta}J(\delta)g(\delta)e^{{\bf{ik}\cdot\delta}}+\frac{J_2}{2}\sum_{\delta'}g(\delta')e^{{\bf{ik}\cdot\delta'}}, 
\label{A1} 
\end{equation} 
and
\begin{eqnarray} 
B_{\bf k}&=&\frac{1}{2}\sum_{\delta}J(\delta)[g(\delta)-f(\delta)(1-e^{{\bf{ik}\cdot\delta}})]\nonumber\\ 
&+&\frac{J_2}{2}\sum_{\delta'}[g(\delta')-f(\delta')(1-e^{{\bf{ik}\cdot\delta'}})]+\mu.
\label{B1} 
\end{eqnarray} 

In equation \eqref{M}, $M_0$ denotes the spontaneous magnetization. In terms of DM bosons, $M_0$ would be  the order parameter of BEC transition, hence  the physical meaning of  $M_0=\langle a_{{\bf k}=0}^\dag a_{\bf k=0}\rangle/N=\langle b_{\bf k=0}^\dag b_{\bf k=0}\rangle/N$, is the number bosons condensed in the zero energy. Therefore, the nonzero value of condensate is an indication of existence of long-range ordering (LRO) in magnetic state of the spin system. 


The magnon energy spectrum is also given by 
\begin{equation} 
\omega_{\bf k}=\sqrt{B_{\bf k}^2-A_{\bf k}^2}, 
\label{magnon1} 
\end{equation} 
At ${\bf k}={\bf 0}$, we have from Eqs. \eqref{A1} and \eqref{B1}  that $B_{{\bf k=0}}=A_{{\bf k=0}}+\mu$. 

For each ordering wave vector ${\bf Q}$, the spin-spin correlations function can  be obtained as

\begin{widetext}
\begin{eqnarray} 
\langle S_i.S_j\rangle&=-\frac{1}{2}[(S+\frac{1}{2}-f(0)+g_{ij})^2(1-\cos({\bf Q}\cdot{\bf r}_{ij}+\phi))-(S+\frac{1}{2}-f(0)+f_{ij})^2(1+\cos({\bf Q}\cdot{\bf r}_{ij}+\phi))],
\label{ss-correlation1} 
\end{eqnarray} 
\end{widetext}
for $\ i\in A\ {\rm and}\ j\in B $ and 
\begin{widetext}
\begin{eqnarray} 
\langle S_i.S_j\rangle&=-\frac{1}{2}[(S+\frac{1}{2}-f(0)+g_{ij})^2(1-\cos({\bf Q}\cdot{\bf r}_{ij}))-(S+\frac{1}{2}-f(0)+f_{ij})^2(1+\cos({\bf Q}\cdot{\bf r}_{ij}))],
\label{ss-correlation2} 
\end{eqnarray} 
\end{widetext}
for $\ i \ {\rm and} \ j \in A \ {\rm or}\ B$.

\section{Derivation of self-consistent equations for N{\'e}el.II and III states} 
\label{app.2}

For N{\'e}el.II and III states, owing to their four-sublattice magnetic pattern (Fig.\ref{ordered}-b,c),  we define the Bogoliubov transformations as 
\begin{widetext}
\begin{eqnarray} 
&&\alpha_{1,{\bf k}}=\frac{1}{\sqrt{2}}(\cosh({\theta^+_{\bf k}})a_{\bf k}-\sinh({\theta^+_{\bf k}})b_{-\bf k}^\dag
-\sinh({\theta^+_{\bf k}})c_{-\bf k}^\dag+\cosh({\theta^+_{\bf k}})d_{\bf k}),\nonumber\\ 
&&\beta_{1,-{\bf k}}^\dag=\frac{1}{\sqrt{2}}(-\sinh({\theta^+_{\bf k}})a_{\bf k}+\cosh({\theta^+_{\bf k}})b_{-\bf k}^\dag
+\cosh({\theta^+_{\bf k}})c_{-\bf k}^\dag-\sinh({\theta^+_{\bf k}})d_{\bf k}),\nonumber\\ 
&&\alpha_{2,-\bf k}^\dag=\frac{1}{\sqrt{2}}(\cosh({\theta^-_{\bf k}})a_{k}-\sinh({\theta^-_{\bf k}})b_{-\bf k}^\dag
+\sinh({\theta^-_{\bf k}})c_{-\bf k}^\dag-\cosh({\theta^-_{\bf k}})d_{\bf k}),\nonumber\\ 
&&\beta_{2,\bf k}=\frac{1}{\sqrt{2}}(-\sinh({\theta^-_{\bf k}})a_{\bf k}+\cosh({\theta^-_{\bf k}})b_{-\bf k}^\dag
-\cosh({\theta^-_{\bf k}})c_{-\bf k}^\dag+\sinh({\theta^-_{\bf k}})d_{\bf k}),
\label{Bogoc} 
\end{eqnarray} 
\end{widetext}
Then following a similar approach discussed in Appendix \ref{app.1},  after
minimizing  the energy with respect to $\theta_{\bf k}^{-}$ and $\theta_{\bf k}^{+} $, that is  
\begin{equation} 
\frac{\partial[E-\mu f(0)]}{\partial \theta_{\bf k}^\pm}=0,\\ 
\label{minc} 
\end{equation} 
and defining $A^{\pm}_{\bf k}$ and $B^{\pm}_{\bf k}$ as

\begin{equation} 
A_{\bf k}^\pm=\frac{1}{2}\sum_{\delta}J(\delta) g(\delta)e^{{\bf{ik}\cdot\delta}}\pm\frac{J_2}{2}\sum_{\delta'}g(\delta')e^{{\bf{ik}\cdot\delta'}},
\label{A2} 
\end{equation} 
and

\begin{eqnarray} 
B_{\bf k}^\pm&=&\frac{1}{2}\sum_{\delta}J(\delta)[g(\delta)-f(\delta)(1\pm e^{{\bf{ik}\cdot\delta}})]\nonumber\\ 
&+&\frac{J_2}{2}\sum_{\delta'}[g(\delta')-f(\delta')(1-e^{{\bf{ik}\cdot\delta'}})]+\mu, 
\label{B2} 
\end{eqnarray} 
one finds the following set of self-consistent equations
\begin{widetext} 
\begin{eqnarray} 
&&M_0=S+\frac{1}{2}-\frac{1}{2N}\sum_{{\bf k}\neq 0}^\prime (\frac{B_{\bf k}^+}{\omega_{\bf k}^+}+\frac{B_{\bf k}^-}{\omega_{\bf k}^-}),\nonumber\\ 
&&f_{ij}=M_0+\frac{1}{2N}\sum_{{\bf k}\neq 0}^\prime(\frac{B_{\bf k}^+}{\omega_{\bf k}^+}-\frac{B_{\bf k}^-}{\omega_{\bf k}^-})e^{i\bf{k}\cdot\bf{r}_{ij}}, \ \ {\rm for}\ ( i\in A ; j\in D) \ {\rm or}\ ( i\in B\ ; j\in C) \nonumber\\ 
&&f_{ij}=M_0+\frac{1}{2N}\sum_{{\bf k}\neq 0}^\prime(\frac{B_{\bf k}^+}{\omega_{\bf k}^+}+\frac{B_{\bf k}^-}{\omega_{\bf k}^-})e^{i\bf{k}\cdot\bf{r}_{ij}},\ \ {\rm for}\ i,j\in A,{\rm }B,{\rm}C,{\rm or\rm} \ D\nonumber\\ 
&&g_{ij}=M_0+\frac{1}{2N}\sum_{{\bf k}\neq 0}^\prime(\frac{A_{\bf k}^+}{\omega_{\bf k}^+}+\frac{A_{\bf k}^-}{\omega_{\bf k}^-})e^{i\bf{k}\cdot\bf{r}_{ij}},\ \ {\rm for}\ ( i\in A ; j\in B) \ {\rm or}\ ( i\in C\ ; j\in D)\nonumber\\ 
&&g_{ij}=M_0+\frac{1}{2N}\sum_{{\bf k}\neq 0}^\prime(\frac{A_{\bf k}^+}{\omega_{\bf k}^+}-\frac{A_{\bf k}^-}{\omega_{\bf k}^-})e^{i\bf{k}\cdot\bf{r}_{ij}},\ \ {\rm for}\ ( i\in A ; j\in C) \ {\rm or}\ ( i\in B\ ; j\in D)\nonumber\\ 
&& {\rm otherwise}  \hspace{0.5cm} f_{ij}=g_{ij}=0. 
\label{fg2} 
\end{eqnarray} 
\end{widetext} 

N{\'e}el.II consists of four sublattices, then there are  two branches of magnon excitations for this phase given by 
\begin{equation} 
\begin{split} 
&\omega_{1,{\bf k}}=\sqrt{{B_{\bf k}^+}^2-{A_{\bf k}^+}^2},\\
&\omega_{2,{\bf k}}=\sqrt{{B_{\bf k}^-}^2-{A_{\bf k}^-}^2}. 
\end{split} 
\label{magnon2} 
\end{equation} 



\begin{thebibliography}{99}

\bibitem{InCuVO} V. Kataev, A. M{\"o}ller, U. L{\"o}w, W. Jung, N. Schittner, M. Kriener, and A. Freimuth, 
                                J. Magn. Magn. Mater. {\bf 310}, 290  (2005).
\bibitem{bmno1} O. Smirnova, M. Azuma, N. Kumada, Y. Kusano, M. Matsuda, Y. Shimakawa, T. Takei, Y. Yonesaki, and N. Kinomura, %
Journal of the American Chemical Society {\bf 131}, 8313  (2009). 
\bibitem{CuNiSbO} J. H. Roudebush, N. H. Andersen, R. Ramlau, V. O. Garlea, R. Toft-Petersen, P. Norby, R. Schneider, J. N.
Hay, and R. J. Cava, Inorg. Chem. {\bf 52}, 6083 (2013).
\bibitem{InCuVO2} M. Yehia, E. Vavilova, A. M{\"o}ller, T. Taetz, U. L{\"o}w, R. Klingeler, V. Kataev, and B. B{\"u}chner,
                                  Phys. Rev. B, \textbf{81}, 060414(R) (2010).
\bibitem{meng} Z. Y. Meng, T. C. Lang, S. Wessel, F. F. Assaad, A. Muramatsu, Nature {\bf 464} 847 (2010).
\bibitem{debate1} S. Sorella, Y. Otsuka, and  S. Yunoki, Scientific Reports {\bf 2}, 992 (2012);
\bibitem{debate2} F.F. Assaad, and I.F. Herbut, Phys. Rev. X {\bf 3}, 031010 (2013).
\bibitem{debate3} B. K. Clark, arXiv:1305.0278.
\bibitem{katsura} S. Katsura, T. Ide, and Y. Morita, J. Stat. Phys. \textbf{42}, 381 (1986).
\bibitem{QMC} J. D. Reger, J. A. Riera, and A. P. Young, J. Phys.: Condens. Matter \textbf{1}, 1855 (1989).
\bibitem{nlsigma} T. Einarsson and H. Johannesson, Phys. Rev. B \textbf{43}, 5867 (1991).
\bibitem{sw} W. H. Zheng, J. Oitmaa and C. J. Hamer Phys. Rev. B {\bf 44}, 10789 (1991).
\bibitem{series} J. Oitmaa, C. J. Hamer and Zheng Weihong, Phys. Rev. B {\bf 45}, 9834 (1992).
\bibitem{fouet} J. B. Fouet, P. Sindzingre, and C. Lhuillier, Eur. Phys. J. B \textbf{20}, 241 (2001).
\bibitem{takano} K. Takano, Phys. Rev. B {\bf 74}, 140402(R) (2006).
\bibitem{noorbakhsh} Z. Noorbakhsh, F. Shahbazi, S. A. Jafari, G. Baskaran, J. Phys. Soc. Jpn. {\bf 78}, 054701 (2009).
\bibitem{kawamura} S. Okumura, H. Kawamura, T. Okubo, and Y. Motome, J. Phys. Soc. Jpn. {\bf 79}, 114705 (2010).
\bibitem{Aron2010} A. Mulder, R. Ganesh, L. Capriotti, and A. Paramekanti, \textbf{81}, 214419 (2010).
\bibitem{sd2} J. Oitmaa, R. R. P. Singh, Phys. Rev. B {\textbf 85}, 014428 (2012).
\bibitem{mosadeq} H. Mosadeq, F. Shahbazi, and S. A. Jafari, J. Phys. Condens. Matter, {\textbf 23}, 226006 (2011).
\bibitem{pvb2} A. F. Albuquerque, D. Schwandt, B. Hetenyi, S. Capponi, M. Mambirini, A. M. Lauchli, Phys. Rev. B {\bf 84}, 024406 (2011).
\bibitem{pvb3} J. Reuther, D. A. Abanin, T. Thomale, Phys. Rev. B {\bf 84}, 014417 (2011).
\bibitem{pvb4} P. H. Y. Li, R. F. Bishop, D. J. J. Farnell,  and C. E. Campbell, J. Phys.: Condens. Matter {\bf 24},  236002 (2012);
Phys. Rev. B {\bf 86}, 144404 (2012).
\bibitem{pvb5} R. F. Bishop, P. H. Y. Li, and C. E. Campbell, J. Phys.:Condens. Matter {\bf 25}, 306002 (2013).
\bibitem{pvb6} R. Ganesh,  S. Nishimoto, and J. van den Brink, Phys. Rev. B {\bf 87}, 054413 (2013).
\bibitem{pvb7} Z. Zhu, D. A. Huse, and S. R. White, Phys. Rev. Lett. {\bf 110}, 127205 (2013).
\bibitem{pvb8} S.-S. Gong, D. N. Sheng, O. I. Motrunich, and M. P. A. Fisher, Phys. Rev. B {\bf 88}, 165138 (2013).
\bibitem{SB-lamas1}   D. C. Cabra, C. A. Lamas, and H. D. Rosales  Phys. Rev. B {\bf 83}, 094506 (2011).
\bibitem{SB-wang} F. Wang, Phys. Rev. B \textbf{82}, 024419 (2010).
\bibitem{SF-lu} Y-M. Lu, and Y. Ran, Phys. Rev. B \textbf{84}, 024420 (2011).
\bibitem{VMC-sondhi} B. K. Clark, D. A. Abanin, and S. L. Sondhi, Phys. Rev. Lett. {\bf 107}, 087204 (2011).
\bibitem{SB-lamas2} H. Zhang, and C. A. Lamas, Phys. Rev. B {\bf 87}, 024415 (2013).
\bibitem{SB-china} X.-L. Yu, D.-Y. Liu, P. Li, and  L.-J. Zou, Physica E {\bf 59}, 41 (2014) 
\bibitem{eps} F. Mezzacapo, and M. Boninsegni, Phys. Rev. B {\bf 85}, 060402 (2012).
\bibitem{ring1} H-Y. Yang, A.F. Albuquerque, S. Capponi, A.M. Lauchli, K.P. Schmidt, New J. Phys. {\bf 14}, 115027 (2012). 
\bibitem{ring2} S. Pujari, K. Damle, and F. Alet, Phys. Rev. Lett {\bf 111}, 087203 (2013). 
\bibitem{zare} M. H Zare, H. Mosadeq, F. Shahbazi and S. A. Jafari, J. Phys.: Condens. Matter {\textbf 26}  456004 (2014).
\bibitem{bishop} P. H. Y. Li, R. F. Bishop, and C. E. Campbell, Phys. Rev. B {\bf 89}, 220408(R). 
\bibitem{Ciolo} A. D. Ciolo, J. Carrasquilla, F. Becca, M. Rigol and V. Galitski, Phys. Rev. B {\bf 89}, 094413 (2013).
\bibitem{DMRG-s1} S.-S. Gong, W. Zhu, and D. N. Sheng, Phys. Rev. B 92, 195110 (2015).
\bibitem{bishop-s1} P. H. Y. Li and R. F. Bishop, arXiv: 1602.08915. 
\bibitem{Takahashi} M. Takahashi, Phys. Rev. B, \textbf{40}, 2494 (1989).
\bibitem{Ting} J. H. Xu and C. S. Ting, Phys. Rev. B, \textbf{43}, 6177 (1991).
\bibitem{dotsen} A. V. Dotsenko and O. P. Sushkov, Phys. Rev. B, \textbf{50}, 13821 (1994).
\bibitem{cirac} P. Hauke, T. Roscilde, V. Murg, J. I. Cirac and R. Schmied, New J. Phys. \textbf{12}, 053036 (2010)

\end{thebibliography}
 \end{document}